\documentclass[aps,floats,twocolumn,prl,nofootinbib]{revtex4-2}
\usepackage{amsfonts,amsmath,amssymb,ascmac,bm,tensor}
\usepackage{fnpct} 
\usepackage{comment}
\usepackage{ifpdf}
\usepackage{slashed}
\usepackage{color}
\usepackage[mathscr]{eucal}
\usepackage[utf8]{inputenc}
\usepackage{physics}
\usepackage{cancel}
\usepackage{soul}
\usepackage{simpler-wick}
\usepackage{hyperref}
\usepackage{appendix}

\def\ddd{\mathrm{d}}
\def\l{\left}
\def\r{\right}
\def\nn{\nonumber}

\begin{document}
\title{Incorporating Backreaction in One-Loop Corrections in Ultra-Slow-Roll Inflation}
\author{Cheng-Jun Fang$^{1,2}$}
\email{fangchengjun@itp.ac.cn}
\author{Zhen-Hong Lyu$^{1,2}$}
\email{lyuzhenhong@itp.ac.cn}
\author{Chao Chen$^{3}$}
\email{cchao012@just.edu.cn}
\author{Zong-Kuan Guo$^{1,2,4}$}
\email{guozk@itp.ac.cn}
	\affiliation{$^1$Institute of Theoretical Physics, Chinese Academy of Sciences, P.O. Box 2735, Beijing 100190, China}
	\affiliation{$^2$School of Physical Sciences, University of Chinese Academy of Sciences, No.19A Yuquan Road, Beijing 100049, China}
   \affiliation{$^3$School of Science, Jiangsu University of Science and Technology, Zhenjiang 212100, China}
   \affiliation{$^4$School of Fundamental Physics and Mathematical Sciences, Hangzhou Institute for Advanced Study, University of Chinese Academy of Sciences, Hangzhou 310024, China}

\begin{abstract} 
\noindent

We investigate the one-loop quantum correction to the primordial curvature power spectrum in the ultra-slow-roll (USR) inflation scenario, incorporating the backreaction effect from curvature perturbations. In the spatially-flat gauge, we expand the background inflaton field up to second order and identify the one-loop level backreaction term in the action. Utilizing a gauge transformation, we derive the comoving curvature interaction Hamiltonian in the presence of the backreaction term and calculate the one-loop correction using the in-in formalism.
Our results reveal that the one-loop super-horizon corrections previously reported in the literature are canceled by the backreaction contributions. This finding underscores the importance of accounting for backreaction effects in the analysis of quantum corrections during USR inflation.

\end{abstract}

\maketitle

\noindent
\emph{{\bf Introduction.}}
The quantum one-loop corrections to the curvature power spectrum have recently posed a challenge. In single-field inflation models with a ultra-slow-roll (USR) period sandwiched between slow-roll (SR) periods, which is a typical scenario for primordial black hole (PBH) production, large amplitude of the small-scale perturbations induce sizeable one-loop correction to the large-scale power spectrum~\cite{Kristiano:2022maq}. 
A direct calculation by in-in formalism yields the one-loop correction in the limit $p \ll k$~\cite{Fumagalli:2023hpa}:
\begin{align}
    \mathcal{P}_\zeta^{\rm loop}(p)
    =c\, \mathcal{P}_\zeta^{0}(p) \int \dd \ln k\ \mathcal{P}_\zeta^{0}(k)+\mathcal{O}\left(\frac{p^3}{k^3}\right) ~,
\end{align}
where $\mathcal{P}_\zeta^{0}$ is the tree-level dimensionless primordial power spectrum, $c$ is a constant. The first term in the expression is a nearly scale-invariant term, while the second term is a volume suppressed term. Reference~\cite{Kristiano:2022maq} was the first to point out that the loop correction $\mathcal{P}_\zeta^{\rm loop}$ can exceed the tree-level power spectrum $\mathcal{P}_\zeta^{0}$ at large scales.

The result is puzzling since curvature perturbations are typically conserved in the super-horizon limit \cite{Bardeen1980,KodamaSasaki1984,Lyth:2004gb,Senatore:2012ya,Assassi:2012et}, sparking debate. Direct in-in calculations often find super-horizon corrections for sudden USR-to-SR transitions \cite{Kristiano:2023scm,Firouzjahi:2023aum,Firouzjahi:2023bkt,Sheikhahmadi:2024peu,kristianoComparingSharpSmooth2024,Choudhury:2023vuj,Franciolini:2023agm,Iacconi:2023ggt,Davies:2023hhn,Maity:2023qzw}, suggesting infrared corrections in non-SR inflation. Different approaches yield contrasting conclusions: consistency relations~\cite{Tada:2023rgp,Kawaguchi:2024rsv,Riotto:2023hoz,Fumagalli:2024absence2}, boundary-term and symmetry-induced quartic Hamiltonian analyses~\cite{Fumagalli:2023hpa,Fumagalli:2024absence2}, and equation-of-motion methods~\cite{Inomata:2024lud} all independently support conservation.
Given this inconsistency, we aim to take the first step toward bridging the gap between direct in-in calculations and other approaches.

Here we emphasize that the backreaction effect, often neglected in previous studies, is critical for in-in calculations of loop-order corrections.
It has long been known that classical perturbations and quantum fluctuations can influence the evolution of the background~\cite{Mukhanov:1996ak,Abramo:1997hu,Schander:2021pgt}, namely the backreaction effect. 
The leading backreaction effect arises from the quadratic terms in perturbations, which can be comparable to the quartic interaction terms and must be included in one-loop calculations. 
Especially in models where perturbations are amplified, such as in USR inflation, the backreaction effect is expected to be sizable, which has been evaluated by lattice simulations ~\cite{Caravano:2024moy,Caravano:2024tlp}.
The backreaction effect in USR inflation has been theoretically studied in Refs.~\cite{Syu:2019uwx,Cheng:2021lif,Cheng:2023ikq}, where the Hartree factorization method was used to numerically solve the equations of motion for perturbations with backreaction. These works equate backreaction effects with quantum loop corrections from short to long modes. 
However, we argue that these effects are distinct, as long-mode perturbations exclude the zero-momentum mode, which corresponds to the background field.
\footnote{Only modes with non-zero momentum can be appropriately canonically quantized analogously to harmonic oscillators.}
In Ref.~\cite{Inomata:2024lud}, curvature perturbations were evaluated using the equation of motion method in the spatially-flat gauge. By including backreactions from perturbations to the background, it was shown that one-loop super-horizon curvature enhancement is entirely canceled, contrasting with results from direct in-in calculations.

In this letter, we propose a systematic method to incorporate backreaction effects within the in-in formalism. Crucially, our method distinguishes between modes with non-zero momentum \( q \), which are quantized canonically as harmonic oscillators, and the \( q = 0 \) modes, which are treated as classical backgrounds. Applying this method to ultra-slow-roll (USR) inflation with a sharp transition, we demonstrate that one-loop super-horizon corrections are precisely canceled by contributions arising from backreaction effects.

\vspace{5pt}
\noindent
\emph{{\bf Backreactions.}}
In order to clarify the role of boundary term in calculating backreaction effects and loop corrections, we need to scrutiny the boundary term from the very beginning. We consider a scalar field $\phi$ minimally coupled with Einstein gravity,
\begin{align} \label{eq:action1}
S &={1 \over 2} \int \ddd^4{x} \sqrt{-g} R + S_{\rm GHY} + S_{\phi} ~,
\nonumber \\
S_{\phi} &=\int \ddd^4{x} \sqrt{-g} \left[ -\frac{1}{2}\partial_\mu \phi \partial^\mu \phi-V(\phi)\right] ~,
\end{align}
where the Gibbons-Hawking-York boundary term $S_{\rm GHY}$ is introduced for the well-defined variation principle~\cite{Gibbons1977} (see Appendix~\ref{appendix:boundary} for details). Adopting the Arnowitt-Deser-Misner (ADM) decomposition of the metric,
\begin{align} \label{eq:adm}
	\dd s^2&=-N^2 \dd t^2+\gamma_{i j}\left(\dd x^i+N^i \dd t\right)\left(\dd x^j+N^j \dd t\right) ~,
\end{align}
where $N$ and $N^i$ are the lapse function and shift vector, respectively. $\gamma_{ij}$ is the reduced metric on the 3D hypersurface. Then, the action~\eqref{eq:action1} becomes
\begin{equation}
S = {1 \over 2} \int\dd[4]{x} \sqrt{-g} R_{\rm ADM} + S_{\phi} ~,
\end{equation}
where $R_{\mathrm{ADM}} \equiv {}^{3}R - K^2 + K_{ij} K^{ij}$ with ${}^{3}R$ the Ricci scalar defined with respect to $\gamma_{ij}$, $K_{ij}$ is the extrinsic curvature and $K \equiv \gamma^{ij} K_{ij}$.

We parametrize the perturbations as
\begin{align}
	N=1+\alpha~, \quad N_i=\partial_i \beta~, \quad \gamma_{ij}=a^2 e^{2\zeta}\delta_{ij} ~.
\end{align}
where vector and tensor perturbations have been neglected. 
For our purposes, we adopt the spatially-flat gauge, where $\zeta=0$ and the scalar field $\phi$ serves as the dynamical variable. We separate the background and fluctuation part of $\phi$ as $\phi=\bar\phi+\delta\phi$.
The scale factor $a$ and Lagrange multipliers $\alpha$, $\beta$ are treated as constrains which can in principle be solved from constraint equations $ \delta_\alpha S=\delta_\beta S=0$. However, these equations are highly nonlinear and challenging to solve. A standard approach is to apply a perturbative expansion and solve these constraints order by order. First, we perform an order expansion of the action with respect to $\delta\phi$
\begin{equation}
    S=S_0+ S_1+ S_2+ S_3+ S_4+\mathcal{O}(\delta\phi^5)~.
\end{equation}
To obtain zero-order background equations, varying $S_1$
\begin{equation}
   \delta_\alpha  S_1=\delta_\beta S_1=\delta_{\delta\phi}  S_1=\mathcal{O}(\delta\phi) ~,
\end{equation}
which gives the Friedmann equation and the Klein-Gordon equation,
\begin{align}
     H^2=\frac{1}{3 }\left[V(\bar{\phi})+\frac{1}{2} \dot{\bar{\phi}}^2\right] ~,\nonumber\\
     \ddot{\bar\phi}+3H\dot{\bar{\phi}}+V_{(1)}(\bar{\phi})=0 ~,
\end{align}
where $V_{(n)}\equiv\ddd^nV/\ddd\phi^n$. With specific initial conditions, these equations give zero-order solutions of $a$ and $\bar{\phi}$ which can be denoted as $a_0$ and $\bar{\phi}_0$. Similarly, by varying $S_2$, we can obtain the linear equation of motion for $\delta\phi$ and the first-order terms of $\alpha$ and $\beta$, which are quite standard procedures. 

Things become different when we vary $S_3$ since $\delta S_3$ are of $\mathcal{O}(\delta\phi^2)$ order and thus have non-zero expectation values which can correct the background equations as
\begin{equation}
    \delta_{\delta\phi}  S_1+\left\langle\delta_{\delta\phi}  S_3\right\rangle=\mathcal{O}(\delta\phi^3)\label{br} ~,
\end{equation}
where $\langle...\rangle$ denotes vacuum expectation value for quantum fluctuations, which contains integrals of bilinear terms over a loop momentum.
It is evident that when considering third-order actions and higher orders, backreaction effects from quantum fluctuations emerge simultaneously. As we will see later, these effects are comparable to one-loop corrections in the USR inflation.

According to Eq.~\eqref{br}, we expand the background inflaton field as
\begin{align}
    \bar{\phi}=\bar{\phi}_{0}+\bar{\phi}_2+\mathcal{O}(\delta\phi^4) ~,
\end{align}
where the second term $\bar{\phi}_2$ specifically encodes the backreaction effect from quantum fluctuations. A similar perturbative correction exists for the Hubble parameter $H$. To quantify this, we first express the scale factor as $a=a_0 e^{\rho_2}+O(\delta\phi^4)$, then the second-order deviation from the zero-order background parameter $H$ can be obtained by $\dot{\rho}_2$. However, through explicit calculation we find that these $\rho_2$-corrections are slow-roll suppressed by $\epsilon \equiv-\dot{H}/H^2$, rendering them subdominant compared to the $\bar\phi_2$-backreaction term.

For simplicity, we neglect slow-roll suppressed terms and retain only the lowest-order term in $\epsilon$, and Eq.~\eqref{br} is simplified to
\begin{equation}
   \ddot{\bar{\phi}}_2+3 {H} \dot{\bar{\phi}}_2+ V_{(2)}(\bar{\phi}_0)\bar{\phi}_2=-\frac{1}{2} V_{(3)}(\bar{\phi}_0)\left\langle\delta \phi^2\right\rangle~.\label{phi2} 
\end{equation}
As we will demonstrate later, the action acquires corrections involving $\bar\phi_2$ due to backreaction effects, resulting in a corresponding correction to the two-point function.

\vspace{5pt}
\noindent
\emph{{\bf One-loop corrections.}}
This section summarizes the one-loop corrections to the large-scale power spectrum in SR-USR-SR inflation model and our setup. In contrast to prior studies, we work in the spatially-flat gauge and derive the correlation functions of curvature perturbations through gauge transformations.
\footnote{We thank Jason Kristiano for pointing out this method, see \url{https://indico.cern.ch/event/1433472/contributions/6136634}.}

In the spatially-flat gauge, the dominant contributions to one-loop corrections are as follows:
\begin{align}
    S=S_3+S_4&=-\int \dd t\dd[3]x\ \left( \frac{1}{6}a^3V_{(3)}\delta \phi^3+\frac{1}{24}  a^4V_{(4)}\delta \phi^4\right) \nonumber\\
    &-\frac{1}{2}\int \dd t\dd[3]x\ a^2 V_{(3)} \bar{\phi}_2 \delta \phi^2 ~,\label{S3S4}
\end{align}
where we have neglected slow-roll suppressed terms. The first line involves cubic and quartic self-interaction terms. The second line represents a backreaction correction term.
Since $\bar{\phi}_2\sim\mathcal{O}(\delta\phi^2)$ as shown in \eqref{phi2}, it is of the same order as the quartic self-interaction term. We will discuss the role of this correction term in the next section.

We define
\begin{equation}
    \bar{\zeta}\equiv-\frac{H \delta \phi}{\dot{\bar{\phi}}},\quad P_{2}\equiv\frac{\dot{\bar{\phi}}_{2}}{\dot{\bar{\phi}}_{0}} ~,
\end{equation}
which are two basic quantities in our setup.\footnote{For simplicity, we do linear gauge transformations, i.e., $\zeta=\bar\zeta$ at the end of inflation. The correlation functions of $\zeta$ can be obtained from those of $\bar\zeta$.} 

Substituting $\delta\phi=-{\dot{\bar\phi}_0(1+P_2)\bar\zeta}/{H}$ into the action \eqref{S3S4}, retaining terms up to $\mathcal{O}(\bar{\zeta}^3)$ and $\mathcal{O}(\bar{\zeta}^4)$ separately, and performing the Legendre transformation, we obtain the third-order and the fourth-order Hamiltonian. The third-order bulk Hamiltonian can be decomposed into two parts by integration by parts,
\begin{equation}
    H^{(3)}_{\mathrm{bulk}}=\frac{1}{6}\int\dd[3]x\ (a^2\epsilon \eta')'\bar{\zeta}^3=H^{(3)}+H_{B} ~,\label{3bulkH}
\end{equation}
where
\begin{equation}
    H^{(3)}=-\frac{1}{2}\int\dd[3]x\ (a^2\epsilon \eta') \bar{\zeta}'\bar{\zeta}^2\label{JK} ~,
\end{equation}
\begin{equation}
    H_{B}=\frac{1}{6}\int\dd[3]x\ (a^2\epsilon \eta'\bar{\zeta}^3)' ~.
\end{equation}
The four-point interaction is written as
\begin{equation}
    H^{(4)}_{\mathrm{bulk}}=\frac{1}{6}\int\dd[3]x\ a^4\epsilon^2 V_{(4)} \bar{\zeta}^4 ~.\label{4bulkH}
\end{equation}

The one-loop corrections from the third-order bulk Hamiltonian \eqref{3bulkH} can be calculated as
\begin{align}
    &\langle\bar{\zeta}_{\mathbf{p}}(\tau_{0}){\bar{\zeta}}_{-\mathbf{p}}(\tau_{0})\rangle_{\bar{\zeta}^3}\nonumber\\
    &=\langle\bar{\zeta}_{\mathbf{p}}(\tau_{0}){\bar{\zeta}}_{-\mathbf{p}}(\tau_{0})\rangle^{(3)}+\langle\bar{\zeta}_{\mathbf{p}}(\tau_{0}){\bar{\zeta}}_{-\mathbf{p}}(\tau_{0})\rangle^{(4\mathrm{I})}~,\label{3c}
\end{align}
where the first term in~\eqref{3c} originates from~\eqref{JK}~\cite{Kristiano:2022maq}, which is widely discussed in previous literature.

In the sharp transition case, $\eta'(\tau)=0$ except for sharp transitions around the begin point of USR period $\tau=\tau_s$ and the end point of USR period $\tau=\tau_s$. We can approximate $\eta$ as $\eta=\Delta\eta\delta(\tau-\tau_e)$ where $\Delta\eta\approx -6$ and neglected contribution from $\tau=\tau_s$ because it is much smaller than that from $\tau=\tau_s$~\cite{Kristiano:2022maq}. Then, one derives~\cite{kristianoComparingSharpSmooth2024} 
\begin{equation}
    \langle\!\langle\bar{\zeta}_{\mathbf{p}}(\tau_{0}){\bar{\zeta}}_{-\mathbf{p}}(\tau_{0})\rangle\!\rangle^{(3)}=\frac{1}{4}(\Delta\eta)^2\abs{{\bar{\zeta}}_p(\tau_0)}^2\int\frac{\dd[3]k}{(2\pi)^3}\abs{{\bar{\zeta}}_{k}(\tau_e)}^2~,\label{JKc}
\end{equation}
where $\langle\!\langle\bar{\zeta}_{\mathbf{p}}\left(\tau_{0}\right) \bar{\zeta}_{-\mathbf{p}}\left(\tau_{0}\right)\rangle\!\rangle$ is defined as
\begin{equation}
    \left\langle{\bar\zeta}_{\mathbf{p}}\left(\tau_{0}\right){\bar\zeta}_{\mathbf{p}^{\prime}}\left(\tau_{0}\right)\right\rangle =(2 \pi)^{3} \delta\left(\mathbf{p}+\mathbf{p}^{\prime}\right)\langle\!\langle\bar{\zeta}_{\mathbf{p}}\left(\tau_{0}\right) \bar{\zeta}_{-\mathbf{p}}\left(\tau_{0}\right)\rangle\!\rangle~.
\end{equation}

We identify the second term in \eqref{3c} as the induced fourth-order corrections. In~\cite{kristianoComparingSharpSmooth2024}, starting from the comving gauge, the one-loop corrections  include a contribution from the quartic-induced Hamiltonian, $H_{4\mathrm{I}}= \frac{1}{16}\int\dd[3]x a^2\epsilon\eta'\zeta^4$, which arises from performing a Legendre transformation on the Lagrangian $\mathcal{L}\sim \zeta'\zeta^2$. 

In our case, since we begin with the $\delta\phi$ gauge, we calculate the full correlator using the third-order bulk Hamiltonian \eqref{3bulkH}, integrating by parts and extracting contributions from \eqref{JK}. The remaining terms represent induced fourth-order corrections (see Appendix~\ref{appendix:Fourth-Order} for details).
\begin{align}
    &\langle\!\langle\bar{\zeta}_{\mathbf{p}}(\tau_{0}){\bar{\zeta}}_{-\mathbf{p}}(\tau_{0})\rangle\!\rangle^{(4\mathrm{I})}\nonumber\\
    &=2\int_{-\infty}^{\tau_{0}}\dd \tau\ a^2(\tau)\epsilon(\tau)\eta'(\tau)\eta'(\tau)\abs{\bar{\zeta}_{p}(\tau_{0})}^2\nonumber\\
    &\hspace{1cm}\times\Im[\bar{\zeta}_{p}(\tau_{0})\bar{\zeta}^\star_{p}(\tau)]\int\frac{\dd[3]k}{(2\pi)^3}\abs{\bar{\zeta}_{k}(\tau)}^2~,\label{4Ic}
\end{align}

The one-loop corrections from the quartic self-interaction Hamiltonian is (see Appendix~\ref{appendix:Fourth-Order} for details)
\begin{align}
    &\langle\!\langle\bar{\zeta}_{\mathbf{p}}(\tau_{0}){\bar{\zeta}}_{-\mathbf{p}}(\tau_{0})\rangle\!\rangle^{(4)}\nonumber\\
    &=8\int_{-\infty}^{\tau_{0}}\dd \tau\ a^4(\tau)\epsilon^2(\tau)V_{(4)}(\bar{\phi}_{0}(\tau))\abs{\bar{\zeta}_{p}(\tau_{0})}^2\nonumber\\
    &\hspace{1cm}\times\Im[\bar{\zeta}_{p}(\tau_{0})\bar{\zeta}^\star_{p}(\tau)]\int\frac{\dd[3]k}{(2\pi)^3}\abs{\bar{\zeta}_{k}(\tau)}^2~.\label{4c}
\end{align}

Remarkably, the induced fourth-order correction, as defined in our formulation, differs from the one-loop order contribution obtained directly from $H_{4\mathrm{I}}$. Specifically, it amounts to precisely 2/3 of the contribution from $H_{4\mathrm{I}}$. This characteristic factor emerges as a direct consequence of our systematic exclusion of zero-momentum modes from $\bar{\zeta}$.

\vspace{5pt}
\noindent
\emph{{\bf Backreaction corrections to correlation functions.}}
We intentionally excluded contributions from a significant class of terms in previous sections. Since $P_2 \bar\zeta$ is a third-order term, substituting $\delta\phi=-{\dot{\bar\phi}_0 (1+P_2)\bar\zeta}/{H}$ into the second-order action introduces fourth-order terms, which must be included in one-loop order evaluations. The second-order action is given by:
\begin{align}
    S_{2}&=\int \dd{t}\dd[3]{x}\, a^3\left[ \frac{1}{2}\dot{\delta \phi}^2-\frac{1}{2}\partial_i \delta\phi\partial^i \delta\phi-\frac{1}{2}V_{(2)}(\bar{\phi}_{0})\delta \phi^2 \right]\nonumber\\
    &+\ldots~,\label{S2}
\end{align}
where $\ldots$ represents terms which are slow-roll suppressed and are neglected in the subsequent analysis. Using $\dot{\bar{\phi}}_{0}=H \sqrt{2 \epsilon}$ and retaining terms up to the fourth order, we obtain
\begin{equation}
    S_{P_2}=\int \dd \tau \dd[3] x\,\left(a^{2} P_{2} \sqrt{2 \epsilon}\bar{\zeta} ( \sqrt{2 \epsilon}\bar{\zeta})^{\prime}\right)^{\prime}+(\dots)\fdv{S_{2}}{\zeta}~,
\end{equation}
where we have applied integration by parts and left a boundary term and a term proportional to the equation of motion which does not contribute to two-point functions. The Hamiltonian is
\begin{equation}
    H_{P_2}=-\int\dd[3]x\,[a^2P_{2}(H\epsilon \eta \bar{\zeta}^2+2\epsilon \bar{\zeta}\bar{\zeta}')]'~.
\end{equation}

This contributes to a correction term in two-point functions,
\begin{align}
    &\langle\bar{\zeta}_{\mathbf{p}}\left(\tau_{0}\right) \bar{\zeta}_{-\mathbf{p}}\left(\tau_{0}\right)\rangle_{P_2}\nonumber\\
    &=2\int_{-\infty}^{\tau_0}\dd\tau'\ \Im\left\langle \bar{\zeta}_{\mathbf{p}}\left(\tau_{0}\right)\bar{\zeta}_{-\mathbf{p}}\left(\tau_{0}\right)\int\frac{\dd[3]k\dd[3]k'}{(2\pi)^6}(2\pi)^3\delta(\mathbf{k}+\mathbf{k}')\right.\nonumber\\
    &\hspace{2cm}\times\left.\left[a^2P_{2}(H\epsilon\eta\bar{\zeta}_{\mathbf{k}}\bar{\zeta}_{\mathbf{k}'}+2\epsilon \bar{\zeta}_{\mathbf{k}}\bar{\zeta}_{\mathbf{k}'}')\right]_{\tau=\tau'}'\right\rangle\nonumber\\
    &=\int\frac{\dd[3]k\dd[3]k'}{(2\pi)^6}(2\pi)^3\delta(\mathbf{k}+\mathbf{k}')\nonumber\\
    &\times[2a^2P_{2}(H\epsilon \eta\Im\langle \bar{\zeta}_{\mathbf{p}}\bar{\zeta}_{-\mathbf{p}}\bar{\zeta}_{\mathbf{k}}\bar{\zeta}_{\mathbf{k}'}\rangle+2\epsilon\Im\langle\bar{\zeta}_{\mathbf{p}}\bar{\zeta}_{-\mathbf{p}}\bar{\zeta}_{\mathbf{k}}\bar{\zeta}'_{\mathbf{k}'}\rangle)]_{\tau=\tau_0}~.
\end{align}

Evaluating it on the time boundary, we found the first term to be purely real, which therefore provides no contribution to the total result, and the second term gives
\begin{equation}
    \langle\!\langle\bar{\zeta}_{\mathbf{p}}\left(\tau_{0}\right) \bar{\zeta}_{-\mathbf{p}}\left(\tau_{0}\right)\rangle\!\rangle_{P_2}=-2 P_{2}(\tau_{0})\abs{\bar{\zeta}_{p}(\tau_{0})}^2~.\label{couter}
\end{equation}
Above we have used
\begin{equation}
    \Im[\bar\zeta_k(\tau){\bar\zeta_k^{\star}}(\tau)']=\frac{1}{4\epsilon(\tau) a(\tau)^2}~.
\end{equation}

We will show that \eqref{couter} cancel the contributions from the one-loop diagrams.

To solve $P_2$, first notice the equation of motion of $\bar{\phi}_0$ and $\bar{\phi}_2$ can be expressed as~\cite{Inomata:2024lud}
\begin{equation}
    \mathcal{N}_{0}\dot{\bar{\phi}}_{0}=0~,
\end{equation}
\begin{align}
    \mathcal{N}_{0}\dot{\bar{\phi}}_{2}=&-\frac{a^2}{2}[V_{(3)}(\bar{\phi}_{0})\ev{\delta \phi^2}^\mathbf{\dot{}}+V_{(4)}(\bar{\phi}_{0})\ev{\delta \phi^2}\dot{\bar{\phi}}_{0} ]\nonumber\\
    &-a^2V_{(3)}(\bar{\phi}_{0})\bar{\phi}_2~.
\end{align}
where $\mathcal{N}_{k}$ is a operator defined as
\begin{equation}
    {\mathcal{N}}_{k} \equiv \frac{\partial^{2}}{\partial \eta^{2}}+2 \mathcal{H} \frac{\partial}{\partial \eta}+k^{2}+a^{2} V_{(2)}(\bar{\phi}_{0})~.
\end{equation}
We immediately observe that the contribution from the term $-a^2V_{(3)}(\bar{\phi}_{0})\bar{\phi}_2$ can be canceled by a term not explicitly mentioned, $H_{\rm br}=\frac{1}{2}\int \dd t\dd[3]x\ 2a^2\epsilon V_{(3)} \bar{\phi}_2 \bar{\zeta}^2$, which is directly induced by backreactions, as seen from the second line of \eqref{S3S4}.

Combing the above two equations, notice that
\begin{equation}
    \ev{\delta \phi^2}^{\mathbf{\dot{}}}=2\epsilon \ev{\bar{\zeta}^2}^{\mathbf{\dot{}}}+2\epsilon H\eta\ev{\bar{\zeta}^2},\quad \ev{\delta\phi^2}=2\epsilon\ev{\bar{\zeta}^2}~.
\end{equation}
and partially express $V_{(3)}/\dot{\bar{\phi}}_{0}=-(a^2\epsilon \eta')'/(4a^4H\epsilon^2)$, we can obtain the equation of motion of $P_{2}$

\begin{equation}
    \mathcal{M}_{0}P_{2}=\frac{(a^2\epsilon \eta')'}{4a^2\epsilon}\left( \frac{\ev{\bar{\zeta}^2}^{\mathbf{\dot{}}}}{H}+\eta\ev{\bar{\zeta}^2} \right)-a^2\epsilon V_{(4)}\ev{\bar\zeta^2}~.\label{eomP2}
\end{equation}
where
\begin{equation}
    \mathcal{M}_{k}\equiv\pdv[2]{\tau}+(2+\eta)aH\pdv{\tau}+k^2~.
\end{equation}
We can calculate $P_2$ by solving \eqref{eomP2} with the Green function method.
\begin{align}
    P_2(\tau)&=\int_{-\infty}^{\tau}\dd\tau' G(\tau;\tau')\left[\frac{(a^2\epsilon \eta')'}{4a^2\epsilon}\left( \frac{\ev{\bar{\zeta}^2}^{\mathbf{\dot{}}}}{H}+\eta\ev{\bar{\zeta}^2} \right)\right]\nonumber\\
    &+\int_{-\infty}^{\tau}\dd\tau' G(\tau;\tau')\left[-a^2\epsilon V_{(4)}\ev{\bar\zeta^2}\right]~.
\end{align}
$G(\tau;\tau')$ is the zero-momentum mode Green function defined as $\mathcal{M}_{0}G(\tau;\tau')=\delta(\tau-\tau')$. 
Notice that $\bar{\zeta}$ satisfies the same equation as $P_2$, except a zero source term
\begin{equation}
    \mathcal{M}_{k}\bar{\zeta}_{\mathbf{k}}=0~,\ k\neq 0~.
\end{equation}
The corresponding Green function is
\begin{align}
    G_{k}(\tau;\tau')&=\frac{2i}{W(\tau')}\Im[\bar{\zeta}_{k}(\tau)\bar{\zeta}_{k}^\star(\tau')]\nonumber\\
    &=-4a^2(\tau')\epsilon(\tau')\Im[\bar{\zeta}_{k}(\tau)\bar{\zeta}_{k}^\star(\tau')]~.
\end{align}
with $W(\tau')=(\bar{\zeta}'\bar{\zeta}^\star-\bar{\zeta}\bar{\zeta}^\star{}')|_{\tau'}$ the Wronskian. Here we have performed operator expansion, $\bar{\zeta}(\mathbf{x},\tau)=\int\frac{\dd[3]k}{(2\pi)^3}\bar{\zeta}_{\mathbf{k}}(\tau)e^{i\mathbf{k}\mathbf{x}}$ and $\bar{\zeta}_{\mathbf{k}}=\bar{\zeta}_{k}a_{\mathbf{k}}+\bar\zeta_{k}^\star a_{-\mathbf{k}}^{\dagger}$.

Note that $\bar{\zeta}$ does not include zero-momentum modes since they are already absorbed the background field $\bar{\phi}$ or $P_2$. Since $\mathcal{M}_k$ is continuous in $k=0$, we can express the zero-mode Green function using non-zero modes $\bar{\zeta}$, i.e. $G(\tau;\tau')=\lim_{q\to 0}G_q(\tau;\tau')$.

We denote $P_{2}^{(1)}$ and $P_{2}^{(2)}$ as the first and the second line of the right-hand side of \eqref{eomP2}, respectively,
\begin{align}
    P_{2}^{(1)}(\tau_{0})&=-\int_{-\infty}^{\tau_0} \dd \tau\ [a^2(\tau)\epsilon(\tau)\eta'(\tau)]'\Im[\bar{\zeta}_{q}(\tau_{0})\bar{\zeta}_{q}^\star(\tau)]\nonumber\\
    &\times\left( \frac{\ev{\bar{\zeta}^2(\tau)}^{\mathbf{\dot{}}}}{H}+\eta(\tau)\ev{\bar{\zeta}^2(\tau)} \right),\ q\to 0~,\label{P21}
\end{align}
\begin{align}
    P_{2}^{(2)}(\tau_{0})&=4\int_{-\infty}^{\tau_{0}} \dd \tau[a^2(\tau)\epsilon(\tau)]^2V_{(4)}(\bar{\phi}_{0}(\tau))\nonumber\\
    &\times \Im[\bar{\zeta}_{q}(\tau_{0})\bar{\zeta}_{q}^\star(\tau)] \ev{\bar{\zeta}^2(\tau)}~,\ q\to 0~.\label{P22}
\end{align}

To solve $P_2^{(1)}$, first notice that
\begin{equation}
    \ev{\bar{\zeta}^2}^{\mathbf{\dot{}}}=\ev{\dot{\bar{\zeta}}\bar{\zeta}+\bar{\zeta}\dot{\bar{\zeta}}}+H\int\frac{\dd k}{k}\dv{\mathcal P^0_{\bar\zeta}(k)}{\ln k}~,\label{zeta2dot}
\end{equation}
the second term originates from time derivatives of the physical cut-offs. Here we take physical cut-offs for both IR and UV limit and thus 
\begin{equation}
    \ev{\bar{\zeta}^2}=\displaystyle\int_{\Lambda_0\frac{a(\tau)}{a(\tau_*)}}^{\Lambda_1\frac{a(\tau)}{a(\tau_*)}}\dd \ln{k}   \, \mathcal{P}_{\bar{\zeta}}^{0}(k)~,\label{cut}
\end{equation}
where $\Lambda_0$ and $\Lambda_1$ corresponds to lower and upper cut-off energy scale respectively and $\tau_*$ is some fiducial time. {Here we adopt the arguments in~\cite{Huenupi:2024ksc} that physical cut-offs do not introduce additional time scales in dS spacetime.} As a result, time derivatives not only act on $\bar\zeta$ but also on the cut-offs. 

Take derivative with respect to coordinate time $t$ on both side of the equation of motion of $\bar\zeta$, we obtain
\begin{equation}
    \mathcal{M}_{k}\dot{\bar{\zeta}}_{\mathbf{k}}=2Hk^2\bar{\zeta}_{\mathbf{k}}-H\eta'\bar{\zeta}_{\mathbf{k}}'~.
\end{equation}
Define $g_{\mathbf{k}}(\tau)\equiv\dfrac{\dd\bar{\zeta}_{\mathbf{k}}}{\dd\ln k}$, it's straight forward to prove that $g_{\mathbf{k}}$ satisfies
\begin{equation}
    \mathcal{M}_{k}g_{\mathbf{k}}=-2k^2\bar{\zeta}_{\mathbf{k}}~.
\end{equation}
hence $\mathcal{M}_{k}\dot{\bar{\zeta}}_{\mathbf{k}}=-H\mathcal{M}_{k}g_{\mathbf{k}}-H\eta'\bar{\zeta}_{\mathbf{k}}'$, we can solve $\dot{\bar{\zeta}}_{\mathbf{k}}$ with the Green function method
\begin{align}\label{zetadot}
    \dot{\bar{\zeta}}_{\mathbf{k}}&=-\int_{-\infty}^\tau \dd{\tau'} G_{k}(\tau,\tau')H\eta'\bar{\zeta}_{\mathbf{k}}'\nonumber\\
    &-Hg_{\mathbf{k}}+C_{\mathbf{k}}H\bar{\zeta}_{k}+D_{\mathbf{k}}H\bar{\zeta}_{k}^\star~,
\end{align}
where $C_\mathbf{k}$ and $D_\mathbf{k}$ are constants that incorporate the creation and annihilation operators (see Appendix~\ref{appendix:P2_detail} for details). Notably, further calculations reveal that only the first line of \eqref{zetadot} contributes to the final result (see Appendix~\ref{appendix:P2_detail} for a detailed derivation). Substituting this into the correlator, we obtain:
\noindent
\begin{align}
\langle\!\langle\bar{\zeta}_{\mathbf{k}}(\tau)\dot{\bar{\zeta}}_{-\mathbf{k}}(\tau)\rangle\!\rangle
&=4\int_{-\infty}^{\tau_0}\dd \tau'\ \epsilon(\tau')a^2(\tau')H(\tau')\eta'(\tau')\nonumber\\
&\times\Im[\bar{\zeta}_{k}(\tau)\bar{\zeta}_{k}^\star(\tau')]\mathrm{Re}[\bar{\zeta}_{k}(\tau){\bar{\zeta}^\star_{k}}{}'(\tau')]~,\label{zetazetadot}
\end{align}

and the same for $\langle\!\langle\dot{\bar{\zeta}}_{\mathbf{k}}(\tau)\bar{\zeta}_{-\mathbf{k}}(\tau)\rangle\!\rangle$. So the final result reads
\begin{align}
    \ev{\bar{\zeta}^2(\tau)}^{\mathbf{\dot{}}}&=8\int_{-\infty}^\tau\dd \tau'\epsilon(\tau')a^2(\tau')H(\tau')\eta'(\tau')\nonumber\\
    &\times\int\frac{\dd[3]k}{(2\pi)^3}\Im[\bar{\zeta}_{k}(\tau)\bar{\zeta}_{k}^\star(\tau')]\mathrm{Re}[\bar{\zeta}_{k}(\tau){\bar{\zeta}^\star_{k}}{}'(\tau')]~.\label{source1}
\end{align}

Next, notice that we can express the second term inside the parentheses of \eqref{P21} as (see Appendix~\ref{appendix:P2_detail} for details)
\begin{align}
&\eta(\tau)\ev{\bar\zeta^2(\tau)}=\int\frac{\dd[3]k}{(2\pi)^3}\int_{-\infty}^\tau \dd \tau'(4a^2(\tau')\epsilon(\tau')\eta'(\tau'))\nonumber\\
&\hspace{1cm}\times(\Re[\bar{\zeta}_{k}(\tau)\bar{\zeta}_{k}^\star(\tau')]\Im[\bar{\zeta}_{k}(\tau)\bar{\zeta}_{k}^\star{}'(\tau')]\nonumber\\
&\hspace{2.1cm}-\Re[\bar{\zeta}_{k}(\tau)\bar{\zeta}_{k}^\star{}'(\tau')]\Im[\bar{\zeta}_{k}(\tau)\bar{\zeta}_{k}^\star(\tau')])~.\label{source2}
\end{align}
Substituting \eqref{source1} and \eqref{source2} into \eqref{P21} and integrate by parts, the final result reads (see Appendix~\ref{appendix:P2_detail} for details)
\begin{widetext}
\begin{equation}
P_{2}^{(1)}(\tau_{0})=\frac{1}{8}(\Delta\eta)^2\int\frac{\dd[3]k}{(2\pi)^3}\abs{\bar{\zeta}_k(\tau_e)}^2+\int_{-\infty}^{\tau_{0}}\dd \tau\ a^2(\tau)\epsilon(\tau)\eta'(\tau)\eta'(\tau)\Im[\bar{\zeta}_{q}(\tau_{0})\bar{\zeta}^\star_{q}(\tau)]\int\frac{\dd[3]k}{(2\pi)^3}\abs{\bar{\zeta}_{k}(\tau)}^2~,\ q\to 0~.
\label{P21result}
\end{equation}
\end{widetext}

We have approximated
\begin{equation}
    \Im[\bar{\zeta}_{q}(\tau_{0})\bar{\zeta}_q^\star{}'(\tau_{e})]\simeq \frac{1}{4a^2(\tau_{e})\epsilon(\tau_{e})}~,
\end{equation}
and used $\displaystyle\int^{\tau_c}_{-\infty}\eta'(\tau)f(\tau)\dd\tau=\mathcal{C}\Delta\eta f(\tau_e)$, where $\mathcal{C}=1,\frac{1}{2},0$ for $\tau_c>\tau_e$, $\tau_c=\tau_e$, $\tau_c<\tau_e$.

Substituting into \eqref{couter}, the first term of \eqref{P21result} introduces a correction to the two-point correlation function
\noindent
\begin{equation}
    \langle\!\langle\bar{\zeta}_{\mathbf{p}}\left(\tau_{0}\right) \bar{\zeta}_{-\mathbf{p}}\left(\tau_{0}\right)\rangle\!\rangle_{P_2}^{(3)}=-\frac{1}{4}(\Delta\eta)^2\abs{\bar{\zeta}_p(\tau_0)}^2\int\frac{\dd[3]k}{(2\pi)^3}\abs{\bar{\zeta}_{k}(\tau_e)}^2~,
\end{equation}
which cancels out with the third-order loop correction \eqref{JKc}.

The second term of \eqref{P21result} contributes
\begin{align}
    &\langle\!\langle\bar{\zeta}_{\mathbf{p}}(\tau_{0}){\bar{\zeta}}_{-\mathbf{p}}(\tau_{0})\rangle\!\rangle^{(4\mathrm{I})}_{P_2} \nonumber\\
    &=-2\int_{-\infty}^{\tau_{0}}\dd \tau\ a^2(\tau)\epsilon(\tau)\eta'(\tau)\eta'(\tau)\abs{\bar{\zeta}_{p}(\tau_{0})}^2 \nonumber\\
    &\times\Im[\bar{\zeta}_{q}(\tau_{0})\bar{\zeta}^\star_{q}(\tau)]\int\frac{\dd[3]k}{(2\pi)^3}\abs{\bar{\zeta}_{k}(\tau)}^2~,\ q\to 0~.\label{4Icc}
\end{align}
Comparing with the induced fourth-order one-loop correction corrections \eqref{4Ic} in the limit $p\ll k$, it cancels out with \eqref{4Icc}.

Finally we solve $P_2^{(2)}$ from Eq.~\eqref{P22}
\begin{align}
P_{2}^{(2)}(\tau_{0})&=4\int_{-\infty}^{\tau_{0}} \dd \tau[a^2(\tau)\epsilon(\tau)]^2V_{(4)}(\bar{\phi}_{0}(\tau))\nonumber\\
&\times \Im[\bar{\zeta}_{q}(\tau_{0})\bar{\zeta}_{q}^\star(\tau)] \int\frac{\dd[3]k}{(2\pi)^3}\abs{\bar{\zeta}_{k}(\tau)}^2~,\ q\to 0~.
\end{align}
Its contribution to the correlation function is given by
\begin{align}
    &\langle\!\langle\bar{\zeta}_{\mathbf{p}}\left(\tau_{0}\right) \bar{\zeta}_{-\mathbf{p}}\left(\tau_{0}\right)\rangle\!\rangle_{P_2}^{(4)} \nonumber\\
    &=-8\int_{-\infty}^{\tau_{0}} \dd \tau[a^2(\tau)\epsilon(\tau)]^2V_{(4)}(\bar{\phi}_{0}(\tau))\abs{\bar{\zeta}_p(\tau_0)}^2 \nonumber\\
    &\times\Im[\bar{\zeta}_{q}(\tau_{0})\bar{\zeta}_{q}^\star(\tau)] \int\frac{\dd[3]k}{(2\pi)^3}\abs{\bar{\zeta}_{k}(\tau)}^2\ q\to 0~.
\end{align}
This cancels out with the loop contribution from the fourth-order bulk Hamiltonian, as shown in \eqref{4c}.

To summarize, the one-loop corrections to the two-point function \eqref{JKc}, \eqref{4Ic} and \eqref{4c} are canceled out with the backreaction correction \eqref{couter}.

\emph{{\bf Discussions.}}
Despite the progress achieved in this research, there are still several areas that necessitate further investigation. When calculating the source term of $P_2$, we employed techniques like equations of motion to express $\langle\delta \phi^2\rangle^{\mathbf{\dot{}}}.$ in a time integral form. However, a more straightforward approach would be to explicitly write the solutions of first-order perturbations and directly compute their time derivatives. This could potentially offer a clearer understanding of the perturbations and backreaction effects. {However, backreaction effect of primordial perturbations itself remains an open question, especially the highly tricky backreaction from quantum degrees. In this paper, we adopted a semiclassical approach to backreaction~\cite{Schander:2021pgt}, leaving further exploration of quantum backreaction to future work. }

Regarding the momentum integral cut-offs, we adopted physical energy scales for both IR and UV limits, as demonstrated in Eq.~\eqref{cut}. This approach yielded relatively concise and physically meaningful results. However, the comparative advantages and limitations of this physical cut-off scheme, particularly in contrast to the conventional method employing comoving cut-offs for the IR limit, remain to be thoroughly investigated. {Recently, the background renormalization of the inflationary field has received more attention. Reference~\cite{Kristiano:2025ajj} highlights that different regularization schemes yield distinct background corrections; however, these corrections ultimately do not affect observables, specifically the two-point correlation function of perturbations.}

In terms of gauge transformation, we only considered the transformation terms crucial at the end of inflation, focusing on $\bar{\zeta}$ rather than $\zeta$ by not accounting for the complete third-order gauge transformation. In future studies, it would be beneficial to initiate calculations directly from the comoving gauge for a more comprehensive and accurate description of curvature perturbations. Furthermore, we simply neglected terms suppressed by the slow-roll parameter $\epsilon$ and higher-order constraints in our calculations. Future work should aim to carefully evaluate their contributions.

Finally, we have yet to approach this problem from the perspective of symmetry, particularly in the context of consistency relations. In future research, it is essential to explore how to preserve the symmetry of the entire theory during perturbative calculations. By addressing these aspects, we can refine the theoretical framework of inflationary cosmology and gain a deeper understanding of the quantum and classical processes that have shaped the universe. 

\emph{{\bf Acknowledgements.}} We thank Jason Kristiano and Keisuke Inomata for insight comments and discussions. This work is supported in part by the National Key Research and Development Program of China (No. 2020YFC2201501), in part by the National Natural Science Foundation of China (No. 12475067 and No. 12235019). CC is supported by National Natural Science Foundation of China (No. 12433002) and Start-up Funds for Doctoral Talents of Jiangsu University of Science and Technology. CC thanks the supports from The Asia Pacific Center for Theoretical Physics, The Center for Theoretical Physics of the Universe at Institute for Basic Science during his visits.

\paragraph*{Notes added}
During the completion of this work, a related study~\cite{Inomata:2025bqw} by Keisuke Inomata independently investigates loop-level conservation of the superhorizon curvature power spectrum, incorporating backreaction effects within the in-in formalism.

\appendix
\begin{widetext}

\section{Gibbons-Hawking-York Boundary Term}
\label{appendix:boundary}

The variation of the Einstein-Hilbert action is calculated as
\begin{align}  \label{eq:action_EH}
\delta S_{\rm EH} 
&= 
{1 \over 2} \int_\mathcal{V} \ddd^4 x \sqrt{-g} \l( R_{\mu\nu} - \frac12 R g_{\mu\nu} \r) \delta g^{\mu\nu} + {1 \over 2} \int_\mathcal{V} \ddd^4 x \sqrt{-g} \nabla_\alpha V^\alpha
\nn\\&=
{1 \over 2} \int_{\mathcal{V}} \ddd^4 x \sqrt{-g} \l( R_{\mu\nu} - \frac12 R g_{\mu\nu} \r) \delta g^{\mu\nu} - {1 \over 2} \int_\mathcal{\partial \mathcal{V}} \ddd^3 y \sqrt{\gamma} V^\alpha n_\alpha ~,
\end{align} 
where $\partial \mathcal{V}$ is the 3D boundary (i.e., the spacelike hypersurface) of the 4D manifold $\mathcal{V}$ and $y$ refers to the coordinate system on $\partial \mathcal{V}$. The boundary term $\nabla_\alpha V^\alpha$ in the first line comes from the variation of the Ricci scalar, namely $\delta R = R_{\mu\nu} \delta g^{\mu \nu} + \nabla_\alpha \l( g^{\mu\nu} \delta\Gamma_{\mu\nu}^\alpha - g^{\mu\alpha} \delta\Gamma_{\nu\mu}^\nu \r)$, which gives $V^\alpha \equiv g^{\mu\nu} \delta\Gamma_{\mu\nu}^\alpha - g^{\mu\alpha} \delta\Gamma_{\nu\mu}^\nu$.
Note that we have used the Stokes' theorem,
\begin{equation}\label{eq:stokes}
\int_\mathcal{V} \ddd^4 x \sqrt{-g} \nabla_\alpha V^\alpha 
= 
- \int_\mathcal{\partial \mathcal{V}} \ddd^3 y \sqrt{\gamma} V^\alpha n_\alpha ~,
\end{equation} 
to replace the integral $\int_\mathcal{V} \ddd^4 x \sqrt{-g} \nabla_\alpha V^\alpha$ in the first line of Eq.~\eqref{eq:action_EH}. Note that the minus sign appears due to the fact that $\partial \mathcal{V}$ is spacelike. $\gamma$ is the reduced metric on $\partial \mathcal{V}$. In order to evaluate $V^\alpha n_\alpha$ in terms of $\partial_\alpha \delta g_{\mu\nu}$, we need to know the expression of $\delta\Gamma$ on the boundary $\partial \mathcal{V}$. Considering $\delta g_{\mu\nu} = 0$ holds everywhere on $\partial \mathcal{V}$, we derive $\delta \Gamma^\lambda_{\mu\nu} \Big|_{\partial \mathcal{V}} =
\frac12 g^{\rho\lambda} 
\l( \partial_\nu \delta g_{\mu\rho} 
+ \partial_\mu \delta g_{\nu\rho} 
- \partial_\rho \delta g_{\mu\nu} \r)$ and $\delta \Gamma_{\mu\nu}^\mu \Big|_{\partial \mathcal{V}}
= \frac12 g^{\alpha\mu} \partial_\mu \delta g_{\alpha\nu}$.
Then, we can calculate $V^\alpha n_\alpha$ as
\begin{align} 
V^\alpha n_\alpha \Big|_{\partial \mathcal{V}}
= n_\alpha (g^{\mu\nu} \delta\Gamma_{\mu\nu}^\alpha - g^{\mu\alpha} \delta\Gamma_{\nu\mu}^\nu)
= - n^\lambda \gamma^{\mu\nu} \partial_{\lambda} \delta g_{\mu\nu} ~,
\end{align} 
where we have used the fact that the tangent derivative of metric variation vanishes on $\partial \mathcal{V}$, namely $\gamma^{\mu\nu} \partial_{\mu} \delta g_{\lambda\nu} = 0$, since $\delta g_{\mu\nu} = 0 \big|_{\partial \mathcal{V}}$. Hence, the boundary integral in Eq.~\eqref{eq:action_EH} gives
\begin{align} 
\int_\mathcal{V} \ddd^4 x \sqrt{-g} \nabla_\alpha V^\alpha 
&= \int_\mathcal{\partial V} \ddd^3 y \sqrt{\gamma} V^\alpha n_\alpha
\nn\\&=
- \int_\mathcal{\partial V} \ddd^3 y \sqrt{\gamma} n^\lambda \gamma^{\mu\nu} \partial_{\lambda} \delta g_{\mu\nu}
\nn\\&=
- 2 \int_\mathcal{\partial V} \ddd^3 y \sqrt{\gamma} \delta K ~,
\end{align} 
where we have used the formula $\delta K = n^\lambda \gamma^{\mu\nu} \partial_{\lambda} \delta g_{\mu\nu}$ in the last step. Then, we can add the GHY term to the EH action to eliminate the boundary term arising from the variation of $R$, and get the correct Einstein field equation without the need to drop the boundary term by hand. 

Applying the ADM decomposition~\eqref{eq:adm}, we have
\begin{align}  \label{eq:adm_R4}
R &= ~{}^{3}R +  K^2 - K_{ij} K^{ij} - 2 R_{\mu\nu} n^\mu n^\nu
\nn\\&= R_{\rm ADM} - 2 \nabla_\alpha (n^\nu \nabla_\nu n^\alpha - n^\alpha K) ~,
\end{align} 
where we have used the relation $R_{\mu\nu} n^\mu n^\nu = \nabla_\alpha (n^\nu \nabla_\nu n^\alpha - n^\alpha \nabla_\nu n^\nu) - K_{\alpha\beta} K^{\alpha\beta} + K^2$ and define
\begin{equation} 
R_{\rm ADM} \equiv ~{}^{3}R - K^2 + K_{ij} K^{ij} ~.
\end{equation} 
Hence, the 4D integral of $R$ in Eq.~\eqref{eq:adm_R4} gives
\begin{align} 
\int_{\mathcal{V}}\ddd^4x \sqrt{-g} R
&= \int_{\mathcal{V}}\ddd t \ddd^3 x N \sqrt{\gamma} R_{\rm ADM} 
- 2 \int_{\mathcal{V}}\ddd^4x \sqrt{-g} \nabla_\alpha (n^\nu \nabla_\nu n^\alpha - n^\alpha K)
\nn\\&=
\int_{\mathcal{V}}\ddd t \ddd^3 x N \sqrt{\gamma} R_{\rm ADM} 
+ 2 \int_\mathcal{\partial \mathcal{V}} \ddd^3 y \sqrt{\gamma} K ~.
\end{align} 
Since $\nabla_c (n^b n^c n_b) = - \nabla_c n^c$, we have $n^b n^c \nabla_c n_b = 0$. It is clearly seen that the second term of the last line of the above expression will exactly cancel the GHY boundary term in the expression, the well-defined gravitation action is given as
\begin{equation} \label{eq:EHaction_bc_adm}
S = {1\over 2} \int_{\mathcal{V}}\ddd t \ddd^3 x~ N \sqrt{\gamma} R_{\rm ADM}
= {1 \over 2} \int_{\mathcal{V}}\ddd t \ddd^3 x~ N \sqrt{\gamma} \l( ~{}^{3}R - K^2 + K_{ij} K^{ij} \r) ~.
\end{equation} 

\section{The Induced Fourth-Order and Quartic Self-Interaction One-Loop Corrections}
\label{appendix:Fourth-Order}

The third-order bulk Hamiltonian is
\begin{equation}
    H^{(3)}_{\mathrm{bulk}}=\int\dd[3]x\ c'(\tau)\bar{\zeta}^3~,
\end{equation}
where $c(\tau)=\frac{1}{6}a^2\epsilon \eta'$.

Its contribution to the one-loop corrections of the two-point function can be calculated using second-order perturbation theory

\begin{equation}
    \left\langle\bar\zeta_{\mathbf p}\bar\zeta_{-\mathbf p}\right\rangle=\left\langle\bar\zeta_{\mathbf p}\bar\zeta_{-\mathbf p}\right\rangle_{(1,1)}+2\Re\left\langle\bar\zeta_{\mathbf p}\bar\zeta_{-\mathbf p}\right\rangle_{(0,2)}~.
\end{equation}

Performing integration by parts,
\begin{align}
    \left\langle\bar\zeta_{\mathbf p}\bar\zeta_{-\mathbf p}\right\rangle_{(1,1)}&=\int\dd[3]x\dd[3]y\int^{\tau_0}_{-\infty}\dd\tau_1\int^{\tau_0}_{-\infty}\dd\tau_2\, c'(\tau_1)c'(\tau_2)\ev{\bar\zeta^3(\mathbf{x},\tau_1)\bar\zeta_{\mathbf{p}}\bar\zeta_{-\mathbf{p}}\bar\zeta^3(\mathbf{y},\tau_2)}\nonumber\\
    &=\int\dd[3]x\dd[3]y\int^{\tau_0}_{-\infty}\dd\tau_1\int^{\tau_0}_{-\infty}\dd\tau_2\,9c(\tau_1)c(\tau_2)\ev{\bar\zeta^2(\mathbf{x},\tau_1)\bar\zeta'(\mathbf{x},\tau_1)\bar\zeta_{\mathbf{p}}\bar\zeta_{-\mathbf{p}}\bar\zeta^2(\mathbf{y},\tau_2)\bar\zeta'(\mathbf{y},\tau_2)}\nonumber\\
    &\hspace{0.3cm}+c^2(\tau_0)\int\dd[3]x\dd[3]y\ev{\bar\zeta^3(\mathbf{x},\tau_0)\bar\zeta_{\mathbf{p}}\bar\zeta_{-\mathbf{p}}\bar\zeta^3(\mathbf{y},\tau_0)}\nonumber\\
    &\hspace{0.3cm}-\int\dd[3]x\dd[3]y\int^{\tau_0}_{-\infty}\dd\tau\, 6c(\tau_0)c(\tau)\Re\ev{\bar\zeta^2(\mathbf{x},\tau)\bar\zeta'(\mathbf{x},\tau)\bar\zeta_{\mathbf{p}}\bar\zeta_{-\mathbf{p}}\bar\zeta^3(\mathbf{y},\tau_0)}~,\label{11}
\end{align}
\begin{align}
     \left\langle\bar\zeta_{\mathbf p}\bar\zeta_{-\mathbf p}\right\rangle_{(0,2)}&=-\int\dd[3]x\dd[3]y\int^{\tau_0}_{-\infty}\dd\tau_1\int^{\tau_1}_{-\infty}\dd\tau_2\, c'(\tau_1)c'(\tau_2)\ev{\bar\zeta_{\mathbf{p}}\bar\zeta_{-\mathbf{p}}\bar\zeta^3(\mathbf{x},\tau_1)\bar\zeta^3(\mathbf{y},\tau_2)}\nonumber\\
     &=-\int\dd[3]x\dd[3]y\int^{\tau_0}_{-\infty}\dd\tau_1\int^{\tau_1}_{-\infty}\dd\tau_2\, 9c(\tau_1)c(\tau_2)\ev{\bar\zeta_{\mathbf{p}}\bar\zeta_{-\mathbf{p}}\bar\zeta^2(\mathbf{x},\tau_1)\bar\zeta'(\mathbf{x},\tau_1)\bar\zeta^2(\mathbf{y},\tau_2)\bar\zeta'(\mathbf{y},\tau_2)}\nonumber\\
     &\hspace{0.3cm}-\frac{1}{2}c^2(\tau_0)\int\dd[3]x\dd[3]y\ev{\bar\zeta_{\mathbf{p}}\bar\zeta_{-\mathbf{p}}\bar\zeta^3(\mathbf{x},\tau_0)\bar\zeta^3(\mathbf{y},\tau_0)}\nonumber\\
     &\hspace{0.3cm}+\int\dd[3]x\dd[3]y\int^{\tau_0}_{-\infty}\dd\tau\, 3c(\tau_0)c(\tau)\ev{\bar\zeta_{\mathbf{p}}\bar\zeta_{-\mathbf{p}}\bar\zeta^3(\mathbf{x},\tau_0)\bar\zeta^2(\mathbf{y},\tau)\bar\zeta'(\mathbf{y},\tau)}\nonumber\\
     &\hspace{0.3cm}+\int\dd[3]x\dd[3]y\int_{-\infty}^{\tau_{0}}\dd \tau\,\frac{3}{2}c(\tau_0)c(\tau)\left\langle\bar{\zeta}_{\mathbf{p}}(\tau_{0}){\bar{\zeta}}_{-\mathbf{p}}(\tau_{0})[\bar{\zeta}^2(\mathbf{x},\tau)\bar{\zeta}'(\mathbf{x},\tau),\bar{\zeta}^3(\mathbf{y},\tau)]\right\rangle~.\label{02}
\end{align}

The first line in both equations collectively yields the first term in~\eqref{3c}, i.e, the contribution from $H^{(3)}=-\frac{1}{2}\int\dd[3]x\ (a^2\epsilon \eta') \bar{\zeta}'\bar{\zeta}^2$. The contributions from the second and third lines in~\eqref{02} precisely cancel out with those in~\eqref{11}. The remaining contribution, originating from the last line of~\eqref{02}, constitutes the induced fourth-order correction. After transforming into the Fourier space, this correction term is explicitly given by
\begin{align}\label{induced4th}
&\langle\bar{\zeta}_{\mathbf{p}}(\tau_{0}){\bar{\zeta}}_{-\mathbf{p}}(\tau_{0})\rangle^{(4\mathrm{I})}=\frac{1}{4}\int_{-\infty}^{\tau_{0}}\dd \tau\ [a^2(\tau)\epsilon(\tau)]^2\eta'(\tau)\eta'(\tau)\int \prod_{i=1}^4\left[\frac{\dd[3]k_{i}}{(2\pi)^3}\right]\int \frac{\dd[3]q}{(2\pi)^3}(2\pi)^3\delta(\mathbf{k}_{1}+\mathbf{k}_{2}+\mathbf{q})(2\pi)^3(\mathbf{k}_{3}+\mathbf{k}_{4}-\mathbf{q})\nonumber\\
&\hspace{3.2cm}\times\Re\left\langle\bar{\zeta}_{\mathbf{p}}(\tau_{0}){\bar{\zeta}}_{-\mathbf{p}}(\tau_{0})\bar{\zeta}_{\mathbf{k}_{1}}(\tau){\bar{\zeta}}_{\mathbf{k}_{2}}(\tau)\bar{\zeta}_{\mathbf{k}_{3}}(\tau){\bar{\zeta}}_{\mathbf{k}_{4}}(\tau)(\bar\zeta_q'(\tau)\bar\zeta_q^\star(\tau)-\bar\zeta_q\bar(\tau)\bar\zeta_q^{\star'}(\tau))\right\rangle~. 
\end{align}

Using $\Im[\bar\zeta_q'\bar\zeta_q^\star]=-1/(4a^2\epsilon)$, Eq.~\eqref{induced4th} reduces to
\begin{align}
&\langle\bar{\zeta}_{\mathbf{p}}(\tau_{0}){\bar{\zeta}}_{-\mathbf{p}}(\tau_{0})\rangle^{(4\mathrm{I})}=\frac{1}{8}\int_{-\infty}^{\tau_{0}}\dd \tau\ a^2(\tau)\epsilon(\tau)\eta'(\tau)\eta'(\tau)\int \prod_{i=1}^4\left[\frac{\dd[3]k_{i}}{(2\pi)^3}\right]\int \frac{\dd[3]q}{(2\pi)^3}(2\pi)^3\delta(\mathbf{k}_{1}+\mathbf{k}_{2}+\mathbf{q})(2\pi)^3(\mathbf{k}_{3}+\mathbf{k}_{4}-\mathbf{q})\nonumber\\
&\hspace{3.2cm}\times\Im\ev{\bar{\zeta}_{\mathbf{p}}(\tau_{0}){\bar{\zeta}}_{-\mathbf{p}}(\tau_{0})\bar{\zeta}_{\mathbf{k}_{1}}(\tau){\bar{\zeta}}_{\mathbf{k}_{2}}(\tau)\bar{\zeta}_{\mathbf{k}_{3}}(\tau){\bar{\zeta}}_{\mathbf{k}_{4}}(\tau)}~.\label{momentumq}
\end{align}

The one-loop corrections from the quartic self-interaction \eqref{4bulkH} is
\begin{align}
    &\langle\bar{\zeta}_{\mathbf{p}}(\tau_{0}){\bar{\zeta}}_{-\mathbf{p}}(\tau_{0})\rangle^{(4)}=\frac{1}{3}\int_{-\infty}^{\tau_{0}}\dd \tau\ a^4(\tau)\epsilon^2(\tau)V_{(4)}(\bar{\phi}_{0}(\tau))\int\prod_{i=1}^4\left[ \frac{\dd[3]k_{i}}{(2\pi)^3} \right](2\pi)^3 \delta(\mathbf{k}_{1}+\mathbf{k}_{2}+\mathbf{k}_{3}+\mathbf{k}_{4})\nonumber\\
    &\hspace{3.2cm}\times\Im\ev{\bar{\zeta}_{\mathbf{p}}(\tau_{0}){\bar{\zeta}}_{-\mathbf{p}}(\tau_{0})\bar{\zeta}_{\mathbf{k}_{1}}(\tau){\bar{\zeta}}_{\mathbf{k}_{2}}(\tau)\bar{\zeta}_{\mathbf{k}_{3}}(\tau){\bar{\zeta}}_{\mathbf{k}_{4}}(\tau)}~.
\end{align}
{It is important to note that all integrals over perturbation modes in the paper implicitly assume \( k \neq 0 \), since the zero modes (\( k=0 \)) have been explicitly separated in advance. }
Performing wick contraction to the above two equations leads to~\eqref{4Ic} and ~\eqref{4c}. 

The induced fourth-order correction derived here differs from the one-loop order contribution obtained directly from $H_{4\mathrm{I}}$, but is precisely 2/3 times the $H_{4\mathrm{I}}$ contribution. This factor arises because we have already removed $q=0$ modes from $\bar{\zeta}$, which is explicitly reflected in the treatment of the momentum $\mathbf{q}$ in~\eqref{momentumq}. Specifically, in contrast to the calculation of the quartic bulk Hamiltonian contribution (the second equation above), where all momenta are integrated out, we have preserved the momentum $\mathbf{q}$ arising from the commutator in~\eqref{induced4th} and refrained from integrating it out. {A more straightforward understanding is that the reduction in the number of contractions fundamentally arises from the fact that
\begin{equation}
    \int_{q\neq0} \frac{d^3q}{(2\pi)^3}\delta^{(3)}(q)=0 ~.
\end{equation}
}
This distinction is crucial as it prevents the two external momenta from contracting with two $\bar\zeta$ fields originating from the same coordinate point, thereby reducing the total number of possible contractions from 12 to 8. This careful treatment of the momentum structure directly leads to the observed 2/3 factor.

\section{Details in Calculating $P_2$} \label{appendix:P2_detail}

\textit{Derivation of Eq.~\eqref{zetazetadot}}

\begin{align}
\dot{\bar{\zeta}}_{\mathbf{k}}&=-\int_{-\infty}^\tau \dd{\tau'} G_{k}(\tau,\tau')H\eta'\bar{\zeta}_{\mathbf{k}}'\nonumber\\
&\hspace{0.3cm}-Hg_{\mathbf{k}}+C_{\mathbf{k}}H\bar{\zeta}_{k}+D_{\mathbf{k}}H\bar{\zeta}_{k}^\star~.
\end{align}
where $C_\mathbf{k}$ and $D_\mathbf{k}$ are some constants, which include the
creation and the annihilation operators. They turn out to be~\cite{Inomata:2024lud} 
\begin{equation}
    C_\mathbf{k}=-\frac{3}{2}a_\mathbf{k}-\pdv{a_\mathbf{k}}{\ln k},\quad D_\mathbf{k}=-\frac{3}{2}a_{-\mathbf{k}}-\pdv{a_{-\mathbf{k}}}{\ln k}~.
\end{equation}

Define the second line of \eqref{zetadot} as $\mathcal{B}_{\mathbf{k}}=-Hg_{\mathbf{k}}+C_{\mathbf{k}}H\bar{\zeta}_{k}+D_{\mathbf{k}}H\bar{\zeta}_{k}$, 
\begin{align}
\langle\!\langle\bar\zeta_{\mathbf{k}}(\tau)\mathcal{B}_{-\mathbf{k}}(\tau)\rangle\!\rangle
&=-\frac{1}{2}H(\tau)\dv{\abs{\bar\zeta_{k}(\tau)}^2}{\ln k}+H(\tau)\bar\zeta_{k}^\star[\bar\zeta_{\mathbf{k}}^{+}(\tau),D_{-\mathbf{k}}(\tau)]\nonumber\\
&=-\frac{1}{2}H(\tau)\left[ \dv{\abs{\bar\zeta_{k}(\tau)}^2}{\ln k}+3\abs{\bar\zeta_{k}(\tau)}^2 \right]~,
\end{align}
where we have used $[\bar{\zeta}_{\mathbf{k}}^+,D_{-\mathbf{k}}]=-\frac{3}{2}[a_\mathbf{k},a^\dagger_\mathbf{k}]=-\frac{3}{2}$. Noticing that $[C_{\mathbf{k}},\bar{\zeta}_{-\mathbf{k}}^-]=-\frac{3}{2}$, we get the same result for the other half $\langle\!\langle\mathcal{B}_{\mathbf{k}}(\tau)\bar\zeta_{-\mathbf{k}}(\tau)\rangle\!\rangle$. 

In total, the contribution to $\ev{\dot{\bar\zeta}\bar\zeta+\bar\zeta\dot{\bar\zeta}}$ from $\mathcal{B}$ is
\begin{align}
\ev{\dot{\bar{\zeta}}\bar{\zeta}+\bar{\zeta}\dot{\bar{\zeta}}}_{\mathcal{B}}&=-H\int\frac{\dd[3]k}{(2\pi)^3}\left[\dv{\abs{\bar{\zeta}_{k}(\tau)}^2}{\ln k}+3\abs{\bar{\zeta}_{k}(\tau)}^2\right]\nonumber\\
&=-H\int\dd\ln k\dv{\mathcal{P}_{\bar{\zeta}}(k)}{\ln k}~.
\end{align}
which cancels out with the second term in \eqref{zeta2dot}. Hence only the first line of \eqref{zetadot} contributes to $\ev{\bar\zeta\dot{\bar\zeta}}$. After straightforward calculation we obtain \eqref{zetazetadot}.

\


\textit{Derivation of Eq.~\eqref{P21result}}

\

Notice the identity
\begin{equation}
    \eta\ev{\bar\zeta^2}=\int\frac{\dd[3] k}{(2\pi)^3}\int_{-\infty}^\tau \dd \tau' [4a^2(\tau')\epsilon(\tau')]\eta'(\tau')\Im[\bar{\zeta}_k(\tau')\bar{\zeta}^\star{}'_k(\tau')]~,
\end{equation}
we have
\begin{align}
\eta\ev{\bar\zeta^2}&=\int\frac{\dd[3]k}{(2\pi)^3}\int_{-\infty}^\tau \dd \tau'(4a^2\epsilon \eta')\Im[\bar{\zeta}_{k}(\tau)\bar{\zeta}_{k}(\tau')\bar{\zeta}_{k}^\star(\tau)\bar{\zeta}_{k}^\star{}'(\tau')] \nonumber\\
&=\int\frac{\dd[3]k}{(2\pi)^3}\int_{-\infty}^\tau \dd \tau'(4a^2\epsilon \eta')(\Re[\bar{\zeta}_{k}(\tau)\bar{\zeta}_{k}^\star(\tau')]\Im[\bar{\zeta}_{k}(\tau)\bar{\zeta}_{k}^\star{}'(\tau')]-\Re[\bar{\zeta}_{k}(\tau)\bar{\zeta}_{k}^\star{}'(\tau')]\Im[\bar{\zeta}_{k}(\tau)\bar{\zeta}_{k}^\star(\tau')])~.
\end{align}

Approximating $H(\tau)\simeq H(\tau')$ in the first term inside
the parentheses of \eqref{P21}, the source term reads
\begin{equation}
    \frac{\ev{\bar{\zeta}^2(\tau)}^{\mathbf{\dot{}}}}{H}+\eta(\tau)\ev{\bar{\zeta}^2(\tau)}=\int\frac{\dd[3]k}{(2\pi)^3}\int_{-\infty}^\tau \dd \tau'(4a^2\epsilon \eta')(\Re[\bar{\zeta}_{k}(\tau)\bar{\zeta}_{k}^\star(\tau')]\Im[\bar{\zeta}_{k}(\tau)\bar{\zeta}_{k}^\star{}'(\tau')]+\Re[\bar{\zeta}_{k}(\tau)\bar{\zeta}_{k}^\star{}'(\tau')]\Im[\bar{\zeta}_{k}(\tau)\bar{\zeta}_{k}^\star(\tau')])~.
\end{equation}
Substituting into \eqref{P21},
\begin{align}
P_{2}^{(1)}(\tau_{0})&=-4\int_{-\infty}^{\tau_0} \dd \tau\ [a^2(\tau)\epsilon(\tau)\eta'(\tau)]'\Im[\bar{\zeta}_{q}(\tau_{0})\bar{\zeta}_{q}^\star(\tau)]\int_{-\infty}^\tau \dd \tau'[a^2(\tau')\epsilon(\tau')\eta'(\tau')] \nonumber\\
&\times\int\frac{\dd[3]k}{(2\pi)^3}(\Re[\bar{\zeta}_{k}(\tau)\bar{\zeta}_{k}^\star(\tau')]\Im[\bar{\zeta}_{k}(\tau)\bar{\zeta}_{k}^\star{}'(\tau')]+\Re[\bar{\zeta}_{k}(\tau)\bar{\zeta}_{k}^\star{}'(\tau')]\Im[\bar{\zeta}_{k}(\tau)\bar{\zeta}_{k}^\star(\tau')])\ q\to 0~.
\end{align}
After performing integration by parts, three terms arise 
\begin{align}
P_{2}^{(1)}(\tau_{0})&=4\int_{-\infty}^{\tau_0} \dd \tau\ [a^2(\tau)\epsilon(\tau)\eta'(\tau)]\Im[\bar{\zeta}_{q}(\tau_{0})\bar{\zeta}_{q}^\star{}'(\tau)]\int_{-\infty}^\tau \dd \tau'[a^2(\tau')\epsilon(\tau')\eta'(\tau')] \nonumber\\
&\times\int\frac{\dd[3]k}{(2\pi)^3}(\Re[\bar{\zeta}_{k}(\tau)\bar{\zeta}_{k}^\star(\tau')]\Im[\bar{\zeta}_{k}(\tau)\bar{\zeta}_{k}^\star{}'(\tau')]+\Re[\bar{\zeta}_{k}(\tau)\bar{\zeta}_{k}^\star{}'(\tau')]\Im[\bar{\zeta}_{k}(\tau)\bar{\zeta}_{k}^\star(\tau')]) \nonumber\\
&+4\int_{-\infty}^{\tau_0} \dd \tau\ [a^2(\tau)\epsilon(\tau)\eta'(\tau)]\Im[\bar{\zeta}_{q}(\tau_{0})\bar{\zeta}_{q}^\star(\tau)]\int_{-\infty}^\tau \dd \tau'[a^2(\tau')\epsilon(\tau')\eta'(\tau')]\int\frac{\dd[3]k}{(2\pi)^3} \nonumber\\
&\times(\Re[\bar{\zeta}{}'_{k}(\tau)\bar{\zeta}_{k}^\star(\tau')]\Im[\bar{\zeta}_{k}(\tau)\bar{\zeta}_{k}^\star{}'(\tau')]+\Re[\bar{\zeta}_{k}(\tau)\bar{\zeta}_{k}^\star{}'(\tau')]\Im[\bar{\zeta}{}'_{k}(\tau)\bar{\zeta}_{k}^\star(\tau')])\nonumber\\
&+4\int_{-\infty}^{\tau_0} \dd \tau\ [a^2(\tau)\epsilon(\tau)\eta'(\tau)]\Im[\bar{\zeta}_{q}(\tau_{0})\bar{\zeta}_{q}^\star(\tau)][a^2(\tau)\epsilon(\tau)\eta'(\tau)]\int\frac{\dd[3]k}{(2\pi)^3}\abs{\bar{\zeta}_{k}(\tau)}^2\Im[\bar{\zeta}_{k}(\tau)\bar{\zeta}_{k}^\star{}'(\tau)]~.
\end{align}
Note that the second term on the right-hand side is proportional to $\Im[\bar\zeta_q(\tau_0)\bar\zeta^\star(\tau_e)]$, which is approximated as $0$ in~\cite{Kristiano:2022maq} to derive \eqref{JKc}. We use the same approximation here, resulting in only the first and third terms remaining. After integrating out the Dirac delta function, the result is

\begin{equation}
    P_{2}^{(1)}(\tau_{0})=\frac{1}{8}(\Delta\eta)^2\int\frac{\dd[3]k}{(2\pi)^3}\abs{\bar{\zeta}_k(\tau_e)}^2+\int_{-\infty}^{\tau_{0}}\dd \tau\ a^2(\tau)\epsilon(\tau)\eta'(\tau)\eta'(\tau)\Im[\bar{\zeta}_{q}(\tau_{0})\bar{\zeta}^\star_{q}(\tau)]\int\frac{\dd[3]k}{(2\pi)^3}\abs{\bar{\zeta}_{k}(\tau)}^2~,\ q\to 0~.
\end{equation}

\end{widetext}

\bibliographystyle{apsrev4-1}
\bibliography{main}

\begin{thebibliography}{34}%
\makeatletter
\providecommand \@ifxundefined [1]{%
 \@ifx{#1\undefined}
}%
\providecommand \@ifnum [1]{%
 \ifnum #1\expandafter \@firstoftwo
 \else \expandafter \@secondoftwo
 \fi
}%
\providecommand \@ifx [1]{%
 \ifx #1\expandafter \@firstoftwo
 \else \expandafter \@secondoftwo
 \fi
}%
\providecommand \natexlab [1]{#1}%
\providecommand \enquote  [1]{``#1''}%
\providecommand \bibnamefont  [1]{#1}%
\providecommand \bibfnamefont [1]{#1}%
\providecommand \citenamefont [1]{#1}%
\providecommand \href@noop [0]{\@secondoftwo}%
\providecommand \href [0]{\begingroup \@sanitize@url \@href}%
\providecommand \@href[1]{\@@startlink{#1}\@@href}%
\providecommand \@@href[1]{\endgroup#1\@@endlink}%
\providecommand \@sanitize@url [0]{\catcode `\\12\catcode `\$12\catcode `\&12\catcode `\#12\catcode `\^12\catcode `\_12\catcode `\%12\relax}%
\providecommand \@@startlink[1]{}%
\providecommand \@@endlink[0]{}%
\providecommand \url  [0]{\begingroup\@sanitize@url \@url }%
\providecommand \@url [1]{\endgroup\@href {#1}{\urlprefix }}%
\providecommand \urlprefix  [0]{URL }%
\providecommand \Eprint [0]{\href }%
\providecommand \doibase [0]{http://dx.doi.org/}%
\providecommand \selectlanguage [0]{\@gobble}%
\providecommand \bibinfo  [0]{\@secondoftwo}%
\providecommand \bibfield  [0]{\@secondoftwo}%
\providecommand \translation [1]{[#1]}%
\providecommand \BibitemOpen [0]{}%
\providecommand \bibitemStop [0]{}%
\providecommand \bibitemNoStop [0]{.\EOS\space}%
\providecommand \EOS [0]{\spacefactor3000\relax}%
\providecommand \BibitemShut  [1]{\csname bibitem#1\endcsname}%
\let\auto@bib@innerbib\@empty
\bibitem [{\citenamefont {Kristiano}\ and\ \citenamefont {Yokoyama}(2024{\natexlab{a}})}]{Kristiano:2022maq}%
  \BibitemOpen
  \bibfield  {author} {\bibinfo {author} {\bibfnamefont {J.}~\bibnamefont {Kristiano}}\ and\ \bibinfo {author} {\bibfnamefont {J.}~\bibnamefont {Yokoyama}},\ }\href {\doibase 10.1103/PhysRevLett.132.221003} {\bibfield  {journal} {\bibinfo  {journal} {Phys. Rev. Lett.}\ }\textbf {\bibinfo {volume} {132}},\ \bibinfo {pages} {221003} (\bibinfo {year} {2024}{\natexlab{a}})},\ \Eprint {http://arxiv.org/abs/2211.03395} {arXiv:2211.03395 [hep-th]} \BibitemShut {NoStop}%
\bibitem [{\citenamefont {Fumagalli}(2023)}]{Fumagalli:2023hpa}%
  \BibitemOpen
  \bibfield  {author} {\bibinfo {author} {\bibfnamefont {J.}~\bibnamefont {Fumagalli}},\ }\href@noop {} {\  (\bibinfo {year} {2023})},\ \Eprint {http://arxiv.org/abs/2305.19263} {arXiv:2305.19263 [astro-ph.CO]} \BibitemShut {NoStop}%
\bibitem [{\citenamefont {Bardeen}(1980)}]{Bardeen1980}%
  \BibitemOpen
  \bibfield  {author} {\bibinfo {author} {\bibfnamefont {J.~M.}\ \bibnamefont {Bardeen}},\ }\href {\doibase 10.1103/PhysRevD.22.1882} {\bibfield  {journal} {\bibinfo  {journal} {Physical Review D}\ }\textbf {\bibinfo {volume} {22}},\ \bibinfo {pages} {1882} (\bibinfo {year} {1980})}\BibitemShut {NoStop}%
\bibitem [{\citenamefont {Kodama}\ and\ \citenamefont {Sasaki}(1984)}]{KodamaSasaki1984}%
  \BibitemOpen
  \bibfield  {author} {\bibinfo {author} {\bibfnamefont {H.}~\bibnamefont {Kodama}}\ and\ \bibinfo {author} {\bibfnamefont {M.}~\bibnamefont {Sasaki}},\ }\href {\doibase 10.1143/PTPS.78.1} {\bibfield  {journal} {\bibinfo  {journal} {Progress of Theoretical Physics Supplement}\ }\textbf {\bibinfo {volume} {78}},\ \bibinfo {pages} {1} (\bibinfo {year} {1984})}\BibitemShut {NoStop}%
\bibitem [{\citenamefont {Lyth}\ \emph {et~al.}(2005)\citenamefont {Lyth}, \citenamefont {Malik},\ and\ \citenamefont {Sasaki}}]{Lyth:2004gb}%
  \BibitemOpen
  \bibfield  {author} {\bibinfo {author} {\bibfnamefont {D.~H.}\ \bibnamefont {Lyth}}, \bibinfo {author} {\bibfnamefont {K.~A.}\ \bibnamefont {Malik}}, \ and\ \bibinfo {author} {\bibfnamefont {M.}~\bibnamefont {Sasaki}},\ }\href {\doibase 10.1088/1475-7516/2005/05/004} {\bibfield  {journal} {\bibinfo  {journal} {JCAP}\ }\textbf {\bibinfo {volume} {05}},\ \bibinfo {pages} {004} (\bibinfo {year} {2005})},\ \Eprint {http://arxiv.org/abs/astro-ph/0411220} {arXiv:astro-ph/0411220} \BibitemShut {NoStop}%
\bibitem [{\citenamefont {Senatore}\ and\ \citenamefont {Zaldarriaga}(2013)}]{Senatore:2012ya}%
  \BibitemOpen
  \bibfield  {author} {\bibinfo {author} {\bibfnamefont {L.}~\bibnamefont {Senatore}}\ and\ \bibinfo {author} {\bibfnamefont {M.}~\bibnamefont {Zaldarriaga}},\ }\href {\doibase 10.1007/JHEP09(2013)148} {\bibfield  {journal} {\bibinfo  {journal} {JHEP}\ }\textbf {\bibinfo {volume} {09}},\ \bibinfo {pages} {148} (\bibinfo {year} {2013})},\ \Eprint {http://arxiv.org/abs/1210.6048} {arXiv:1210.6048 [hep-th]} \BibitemShut {NoStop}%
\bibitem [{\citenamefont {Assassi}\ \emph {et~al.}(2013)\citenamefont {Assassi}, \citenamefont {Baumann},\ and\ \citenamefont {Green}}]{Assassi:2012et}%
  \BibitemOpen
  \bibfield  {author} {\bibinfo {author} {\bibfnamefont {V.}~\bibnamefont {Assassi}}, \bibinfo {author} {\bibfnamefont {D.}~\bibnamefont {Baumann}}, \ and\ \bibinfo {author} {\bibfnamefont {D.}~\bibnamefont {Green}},\ }\href {\doibase 10.1007/JHEP02(2013)151} {\bibfield  {journal} {\bibinfo  {journal} {JHEP}\ }\textbf {\bibinfo {volume} {02}},\ \bibinfo {pages} {151} (\bibinfo {year} {2013})},\ \Eprint {http://arxiv.org/abs/1210.7792} {arXiv:1210.7792 [hep-th]} \BibitemShut {NoStop}%
\bibitem [{\citenamefont {Kristiano}\ and\ \citenamefont {Yokoyama}(2024{\natexlab{b}})}]{Kristiano:2023scm}%
  \BibitemOpen
  \bibfield  {author} {\bibinfo {author} {\bibfnamefont {J.}~\bibnamefont {Kristiano}}\ and\ \bibinfo {author} {\bibfnamefont {J.}~\bibnamefont {Yokoyama}},\ }\href {\doibase 10.1103/PhysRevD.109.103541} {\bibfield  {journal} {\bibinfo  {journal} {Phys. Rev. D}\ }\textbf {\bibinfo {volume} {109}},\ \bibinfo {pages} {103541} (\bibinfo {year} {2024}{\natexlab{b}})},\ \Eprint {http://arxiv.org/abs/2303.00341} {arXiv:2303.00341 [hep-th]} \BibitemShut {NoStop}%
\bibitem [{\citenamefont {Firouzjahi}(2023)}]{Firouzjahi:2023aum}%
  \BibitemOpen
  \bibfield  {author} {\bibinfo {author} {\bibfnamefont {H.}~\bibnamefont {Firouzjahi}},\ }\href {\doibase 10.1088/1475-7516/2023/10/006} {\bibfield  {journal} {\bibinfo  {journal} {JCAP}\ }\textbf {\bibinfo {volume} {10}},\ \bibinfo {pages} {006} (\bibinfo {year} {2023})},\ \Eprint {http://arxiv.org/abs/2303.12025} {arXiv:2303.12025 [astro-ph.CO]} \BibitemShut {NoStop}%
\bibitem [{\citenamefont {Firouzjahi}(2024)}]{Firouzjahi:2023bkt}%
  \BibitemOpen
  \bibfield  {author} {\bibinfo {author} {\bibfnamefont {H.}~\bibnamefont {Firouzjahi}},\ }\href {\doibase 10.1103/PhysRevD.109.043514} {\bibfield  {journal} {\bibinfo  {journal} {Phys. Rev. D}\ }\textbf {\bibinfo {volume} {109}},\ \bibinfo {pages} {043514} (\bibinfo {year} {2024})},\ \Eprint {http://arxiv.org/abs/2311.04080} {arXiv:2311.04080 [astro-ph.CO]} \BibitemShut {NoStop}%
\bibitem [{\citenamefont {Sheikhahmadi}\ and\ \citenamefont {Nassiri-Rad}(2024)}]{Sheikhahmadi:2024peu}%
  \BibitemOpen
  \bibfield  {author} {\bibinfo {author} {\bibfnamefont {H.}~\bibnamefont {Sheikhahmadi}}\ and\ \bibinfo {author} {\bibfnamefont {A.}~\bibnamefont {Nassiri-Rad}},\ }\href@noop {} {\  (\bibinfo {year} {2024})},\ \Eprint {http://arxiv.org/abs/2411.18525} {arXiv:2411.18525 [astro-ph.CO]} \BibitemShut {NoStop}%
\bibitem [{\citenamefont {Kristiano}\ and\ \citenamefont {Yokoyama}(2024{\natexlab{c}})}]{kristianoComparingSharpSmooth2024}%
  \BibitemOpen
  \bibfield  {author} {\bibinfo {author} {\bibfnamefont {J.}~\bibnamefont {Kristiano}}\ and\ \bibinfo {author} {\bibfnamefont {J.}~\bibnamefont {Yokoyama}},\ }\href {\doibase 10.1088/1475-7516/2024/10/036} {\bibfield  {journal} {\bibinfo  {journal} {JCAP}\ }\textbf {\bibinfo {volume} {10}},\ \bibinfo {pages} {036} (\bibinfo {year} {2024}{\natexlab{c}})},\ \Eprint {http://arxiv.org/abs/2405.12145} {arXiv:2405.12145} \BibitemShut {NoStop}%
\bibitem [{\citenamefont {Choudhury}\ \emph {et~al.}(2024)\citenamefont {Choudhury}, \citenamefont {Gangopadhyay},\ and\ \citenamefont {Sami}}]{Choudhury:2023vuj}%
  \BibitemOpen
  \bibfield  {author} {\bibinfo {author} {\bibfnamefont {S.}~\bibnamefont {Choudhury}}, \bibinfo {author} {\bibfnamefont {M.~R.}\ \bibnamefont {Gangopadhyay}}, \ and\ \bibinfo {author} {\bibfnamefont {M.}~\bibnamefont {Sami}},\ }\href {\doibase 10.1140/epjc/s10052-024-13218-2} {\bibfield  {journal} {\bibinfo  {journal} {Eur. Phys. J. C}\ }\textbf {\bibinfo {volume} {84}},\ \bibinfo {pages} {884} (\bibinfo {year} {2024})},\ \Eprint {http://arxiv.org/abs/2301.10000} {arXiv:2301.10000 [astro-ph.CO]} \BibitemShut {NoStop}%
\bibitem [{\citenamefont {Franciolini}\ \emph {et~al.}(2024)\citenamefont {Franciolini}, \citenamefont {Iovino}, \citenamefont {Taoso},\ and\ \citenamefont {Urbano}}]{Franciolini:2023agm}%
  \BibitemOpen
  \bibfield  {author} {\bibinfo {author} {\bibfnamefont {G.}~\bibnamefont {Franciolini}}, \bibinfo {author} {\bibfnamefont {A.}~\bibnamefont {Iovino}, \bibfnamefont {Junior.}}, \bibinfo {author} {\bibfnamefont {M.}~\bibnamefont {Taoso}}, \ and\ \bibinfo {author} {\bibfnamefont {A.}~\bibnamefont {Urbano}},\ }\href {\doibase 10.1103/PhysRevD.109.123550} {\bibfield  {journal} {\bibinfo  {journal} {Phys. Rev. D}\ }\textbf {\bibinfo {volume} {109}},\ \bibinfo {pages} {123550} (\bibinfo {year} {2024})},\ \Eprint {http://arxiv.org/abs/2305.03491} {arXiv:2305.03491 [astro-ph.CO]} \BibitemShut {NoStop}%
\bibitem [{\citenamefont {Iacconi}\ \emph {et~al.}(2024)\citenamefont {Iacconi}, \citenamefont {Mulryne},\ and\ \citenamefont {Seery}}]{Iacconi:2023ggt}%
  \BibitemOpen
  \bibfield  {author} {\bibinfo {author} {\bibfnamefont {L.}~\bibnamefont {Iacconi}}, \bibinfo {author} {\bibfnamefont {D.}~\bibnamefont {Mulryne}}, \ and\ \bibinfo {author} {\bibfnamefont {D.}~\bibnamefont {Seery}},\ }\href {\doibase 10.1088/1475-7516/2024/06/062} {\bibfield  {journal} {\bibinfo  {journal} {JCAP}\ }\textbf {\bibinfo {volume} {06}},\ \bibinfo {pages} {062} (\bibinfo {year} {2024})},\ \Eprint {http://arxiv.org/abs/2312.12424} {arXiv:2312.12424 [astro-ph.CO]} \BibitemShut {NoStop}%
\bibitem [{\citenamefont {Davies}\ \emph {et~al.}(2024)\citenamefont {Davies}, \citenamefont {Iacconi},\ and\ \citenamefont {Mulryne}}]{Davies:2023hhn}%
  \BibitemOpen
  \bibfield  {author} {\bibinfo {author} {\bibfnamefont {M.~W.}\ \bibnamefont {Davies}}, \bibinfo {author} {\bibfnamefont {L.}~\bibnamefont {Iacconi}}, \ and\ \bibinfo {author} {\bibfnamefont {D.~J.}\ \bibnamefont {Mulryne}},\ }\href {\doibase 10.1088/1475-7516/2024/04/050} {\bibfield  {journal} {\bibinfo  {journal} {JCAP}\ }\textbf {\bibinfo {volume} {04}},\ \bibinfo {pages} {050} (\bibinfo {year} {2024})},\ \Eprint {http://arxiv.org/abs/2312.05694} {arXiv:2312.05694 [astro-ph.CO]} \BibitemShut {NoStop}%
\bibitem [{\citenamefont {Maity}\ \emph {et~al.}(2024)\citenamefont {Maity}, \citenamefont {Ragavendra}, \citenamefont {Sethi},\ and\ \citenamefont {Sriramkumar}}]{Maity:2023qzw}%
  \BibitemOpen
  \bibfield  {author} {\bibinfo {author} {\bibfnamefont {S.}~\bibnamefont {Maity}}, \bibinfo {author} {\bibfnamefont {H.~V.}\ \bibnamefont {Ragavendra}}, \bibinfo {author} {\bibfnamefont {S.~K.}\ \bibnamefont {Sethi}}, \ and\ \bibinfo {author} {\bibfnamefont {L.}~\bibnamefont {Sriramkumar}},\ }\href {\doibase 10.1088/1475-7516/2024/05/046} {\bibfield  {journal} {\bibinfo  {journal} {JCAP}\ }\textbf {\bibinfo {volume} {05}},\ \bibinfo {pages} {046} (\bibinfo {year} {2024})},\ \Eprint {http://arxiv.org/abs/2307.13636} {arXiv:2307.13636 [astro-ph.CO]} \BibitemShut {NoStop}%
\bibitem [{\citenamefont {Tada}\ \emph {et~al.}(2024)\citenamefont {Tada}, \citenamefont {Terada},\ and\ \citenamefont {Tokuda}}]{Tada:2023rgp}%
  \BibitemOpen
  \bibfield  {author} {\bibinfo {author} {\bibfnamefont {Y.}~\bibnamefont {Tada}}, \bibinfo {author} {\bibfnamefont {T.}~\bibnamefont {Terada}}, \ and\ \bibinfo {author} {\bibfnamefont {J.}~\bibnamefont {Tokuda}},\ }\href {\doibase 10.1007/JHEP01(2024)105} {\bibfield  {journal} {\bibinfo  {journal} {JHEP}\ }\textbf {\bibinfo {volume} {01}},\ \bibinfo {pages} {105} (\bibinfo {year} {2024})},\ \Eprint {http://arxiv.org/abs/2308.04732} {arXiv:2308.04732 [hep-th]} \BibitemShut {NoStop}%
\bibitem [{\citenamefont {Kawaguchi}\ \emph {et~al.}(2024)\citenamefont {Kawaguchi}, \citenamefont {Tsujikawa},\ and\ \citenamefont {Yamada}}]{Kawaguchi:2024rsv}%
  \BibitemOpen
  \bibfield  {author} {\bibinfo {author} {\bibfnamefont {R.}~\bibnamefont {Kawaguchi}}, \bibinfo {author} {\bibfnamefont {S.}~\bibnamefont {Tsujikawa}}, \ and\ \bibinfo {author} {\bibfnamefont {Y.}~\bibnamefont {Yamada}},\ }\href {\doibase 10.1007/JHEP12(2024)095} {\bibfield  {journal} {\bibinfo  {journal} {JHEP}\ }\textbf {\bibinfo {volume} {12}},\ \bibinfo {pages} {095} (\bibinfo {year} {2024})},\ \Eprint {http://arxiv.org/abs/2407.19742} {arXiv:2407.19742 [hep-th]} \BibitemShut {NoStop}%
\bibitem [{\citenamefont {Riotto}(2023)}]{Riotto:2023hoz}%
  \BibitemOpen
  \bibfield  {author} {\bibinfo {author} {\bibfnamefont {A.}~\bibnamefont {Riotto}},\ }\href@noop {} {\  (\bibinfo {year} {2023})},\ \Eprint {http://arxiv.org/abs/2301.00599} {arXiv:2301.00599 [astro-ph.CO]} \BibitemShut {NoStop}%
\bibitem [{\citenamefont {Fumagalli}(2024)}]{Fumagalli:2024absence2}%
  \BibitemOpen
  \bibfield  {author} {\bibinfo {author} {\bibfnamefont {J.}~\bibnamefont {Fumagalli}},\ }\href@noop {} {\  (\bibinfo {year} {2024})},\ \Eprint {http://arxiv.org/abs/2408.08296} {arXiv:2408.08296 [astro-ph.CO]} \BibitemShut {NoStop}%
\bibitem [{\citenamefont {Inomata}(2024)}]{Inomata:2024lud}%
  \BibitemOpen
  \bibfield  {author} {\bibinfo {author} {\bibfnamefont {K.}~\bibnamefont {Inomata}},\ }\href {\doibase 10.1103/PhysRevLett.133.141001} {\bibfield  {journal} {\bibinfo  {journal} {Phys. Rev. Lett.}\ }\textbf {\bibinfo {volume} {133}},\ \bibinfo {pages} {141001} (\bibinfo {year} {2024})},\ \Eprint {http://arxiv.org/abs/2403.04682} {arXiv:2403.04682 [astro-ph.CO]} \BibitemShut {NoStop}%
\bibitem [{\citenamefont {Mukhanov}\ \emph {et~al.}(1997)\citenamefont {Mukhanov}, \citenamefont {Abramo},\ and\ \citenamefont {Brandenberger}}]{Mukhanov:1996ak}%
  \BibitemOpen
  \bibfield  {author} {\bibinfo {author} {\bibfnamefont {V.~F.}\ \bibnamefont {Mukhanov}}, \bibinfo {author} {\bibfnamefont {L.~R.~W.}\ \bibnamefont {Abramo}}, \ and\ \bibinfo {author} {\bibfnamefont {R.~H.}\ \bibnamefont {Brandenberger}},\ }\href {\doibase 10.1103/PhysRevLett.78.1624} {\bibfield  {journal} {\bibinfo  {journal} {Phys. Rev. Lett.}\ }\textbf {\bibinfo {volume} {78}},\ \bibinfo {pages} {1624} (\bibinfo {year} {1997})},\ \Eprint {http://arxiv.org/abs/gr-qc/9609026} {arXiv:gr-qc/9609026} \BibitemShut {NoStop}%
\bibitem [{\citenamefont {Abramo}\ \emph {et~al.}(1997)\citenamefont {Abramo}, \citenamefont {Brandenberger},\ and\ \citenamefont {Mukhanov}}]{Abramo:1997hu}%
  \BibitemOpen
  \bibfield  {author} {\bibinfo {author} {\bibfnamefont {L.~R.~W.}\ \bibnamefont {Abramo}}, \bibinfo {author} {\bibfnamefont {R.~H.}\ \bibnamefont {Brandenberger}}, \ and\ \bibinfo {author} {\bibfnamefont {V.~F.}\ \bibnamefont {Mukhanov}},\ }\href {\doibase 10.1103/PhysRevD.56.3248} {\bibfield  {journal} {\bibinfo  {journal} {Phys. Rev. D}\ }\textbf {\bibinfo {volume} {56}},\ \bibinfo {pages} {3248} (\bibinfo {year} {1997})},\ \Eprint {http://arxiv.org/abs/gr-qc/9704037} {arXiv:gr-qc/9704037} \BibitemShut {NoStop}%
\bibitem [{\citenamefont {Schander}\ and\ \citenamefont {Thiemann}(2021)}]{Schander:2021pgt}%
  \BibitemOpen
  \bibfield  {author} {\bibinfo {author} {\bibfnamefont {S.}~\bibnamefont {Schander}}\ and\ \bibinfo {author} {\bibfnamefont {T.}~\bibnamefont {Thiemann}},\ }\href {\doibase 10.3389/fspas.2021.692198} {\bibfield  {journal} {\bibinfo  {journal} {Front. Astron. Space Sci.}\ }\textbf {\bibinfo {volume} {0}},\ \bibinfo {pages} {113} (\bibinfo {year} {2021})},\ \Eprint {http://arxiv.org/abs/2106.06043} {arXiv:2106.06043 [gr-qc]} \BibitemShut {NoStop}%
\bibitem [{\citenamefont {Caravano}\ \emph {et~al.}(2024{\natexlab{a}})\citenamefont {Caravano}, \citenamefont {Franciolini},\ and\ \citenamefont {Renaux-Petel}}]{Caravano:2024moy}%
  \BibitemOpen
  \bibfield  {author} {\bibinfo {author} {\bibfnamefont {A.}~\bibnamefont {Caravano}}, \bibinfo {author} {\bibfnamefont {G.}~\bibnamefont {Franciolini}}, \ and\ \bibinfo {author} {\bibfnamefont {S.}~\bibnamefont {Renaux-Petel}},\ }\href@noop {} {\  (\bibinfo {year} {2024}{\natexlab{a}})},\ \Eprint {http://arxiv.org/abs/2410.23942} {arXiv:2410.23942 [astro-ph.CO]} \BibitemShut {NoStop}%
\bibitem [{\citenamefont {Caravano}\ \emph {et~al.}(2024{\natexlab{b}})\citenamefont {Caravano}, \citenamefont {Inomata},\ and\ \citenamefont {Renaux-Petel}}]{Caravano:2024tlp}%
  \BibitemOpen
  \bibfield  {author} {\bibinfo {author} {\bibfnamefont {A.}~\bibnamefont {Caravano}}, \bibinfo {author} {\bibfnamefont {K.}~\bibnamefont {Inomata}}, \ and\ \bibinfo {author} {\bibfnamefont {S.}~\bibnamefont {Renaux-Petel}},\ }\href {\doibase 10.1103/PhysRevLett.133.151001} {\bibfield  {journal} {\bibinfo  {journal} {Phys. Rev. Lett.}\ }\textbf {\bibinfo {volume} {133}},\ \bibinfo {pages} {151001} (\bibinfo {year} {2024}{\natexlab{b}})},\ \Eprint {http://arxiv.org/abs/2403.12811} {arXiv:2403.12811 [astro-ph.CO]} \BibitemShut {NoStop}%
\bibitem [{\citenamefont {Syu}\ \emph {et~al.}(2020)\citenamefont {Syu}, \citenamefont {Lee},\ and\ \citenamefont {Ng}}]{Syu:2019uwx}%
  \BibitemOpen
  \bibfield  {author} {\bibinfo {author} {\bibfnamefont {W.-C.}\ \bibnamefont {Syu}}, \bibinfo {author} {\bibfnamefont {D.-S.}\ \bibnamefont {Lee}}, \ and\ \bibinfo {author} {\bibfnamefont {K.-W.}\ \bibnamefont {Ng}},\ }\href {\doibase 10.1103/PhysRevD.101.025013} {\bibfield  {journal} {\bibinfo  {journal} {Phys. Rev. D}\ }\textbf {\bibinfo {volume} {101}},\ \bibinfo {pages} {025013} (\bibinfo {year} {2020})},\ \Eprint {http://arxiv.org/abs/1907.13089} {arXiv:1907.13089 [gr-qc]} \BibitemShut {NoStop}%
\bibitem [{\citenamefont {Cheng}\ \emph {et~al.}(2022)\citenamefont {Cheng}, \citenamefont {Lee},\ and\ \citenamefont {Ng}}]{Cheng:2021lif}%
  \BibitemOpen
  \bibfield  {author} {\bibinfo {author} {\bibfnamefont {S.-L.}\ \bibnamefont {Cheng}}, \bibinfo {author} {\bibfnamefont {D.-S.}\ \bibnamefont {Lee}}, \ and\ \bibinfo {author} {\bibfnamefont {K.-W.}\ \bibnamefont {Ng}},\ }\href {\doibase 10.1016/j.physletb.2022.136956} {\bibfield  {journal} {\bibinfo  {journal} {Phys. Lett. B}\ }\textbf {\bibinfo {volume} {827}},\ \bibinfo {pages} {136956} (\bibinfo {year} {2022})},\ \Eprint {http://arxiv.org/abs/2106.09275} {arXiv:2106.09275 [astro-ph.CO]} \BibitemShut {NoStop}%
\bibitem [{\citenamefont {Cheng}\ \emph {et~al.}(2024)\citenamefont {Cheng}, \citenamefont {Lee},\ and\ \citenamefont {Ng}}]{Cheng:2023ikq}%
  \BibitemOpen
  \bibfield  {author} {\bibinfo {author} {\bibfnamefont {S.-L.}\ \bibnamefont {Cheng}}, \bibinfo {author} {\bibfnamefont {D.-S.}\ \bibnamefont {Lee}}, \ and\ \bibinfo {author} {\bibfnamefont {K.-W.}\ \bibnamefont {Ng}},\ }\href {\doibase 10.1088/1475-7516/2024/03/008} {\bibfield  {journal} {\bibinfo  {journal} {JCAP}\ }\textbf {\bibinfo {volume} {03}},\ \bibinfo {pages} {008} (\bibinfo {year} {2024})},\ \Eprint {http://arxiv.org/abs/2305.16810} {arXiv:2305.16810 [astro-ph.CO]} \BibitemShut {NoStop}%
\bibitem [{\citenamefont {Gibbons}\ and\ \citenamefont {Hawking}(1977)}]{Gibbons1977}%
  \BibitemOpen
  \bibfield  {author} {\bibinfo {author} {\bibfnamefont {G.~W.}\ \bibnamefont {Gibbons}}\ and\ \bibinfo {author} {\bibfnamefont {S.~W.}\ \bibnamefont {Hawking}},\ }\href {\doibase 10.1103/PhysRevD.15.2738} {\bibfield  {journal} {\bibinfo  {journal} {Physical Review D}\ }\textbf {\bibinfo {volume} {15}},\ \bibinfo {pages} {2738} (\bibinfo {year} {1977})}\BibitemShut {NoStop}%
\bibitem [{\citenamefont {Huenupi}\ \emph {et~al.}(2024)\citenamefont {Huenupi}, \citenamefont {Hughes}, \citenamefont {Palma},\ and\ \citenamefont {Sypsas}}]{Huenupi:2024ksc}%
  \BibitemOpen
  \bibfield  {author} {\bibinfo {author} {\bibfnamefont {J.}~\bibnamefont {Huenupi}}, \bibinfo {author} {\bibfnamefont {E.}~\bibnamefont {Hughes}}, \bibinfo {author} {\bibfnamefont {G.~A.}\ \bibnamefont {Palma}}, \ and\ \bibinfo {author} {\bibfnamefont {S.}~\bibnamefont {Sypsas}},\ }\href {\doibase 10.1103/PhysRevD.110.123536} {\bibfield  {journal} {\bibinfo  {journal} {Phys. Rev. D}\ }\textbf {\bibinfo {volume} {110}},\ \bibinfo {pages} {123536} (\bibinfo {year} {2024})},\ \Eprint {http://arxiv.org/abs/2406.07610} {arXiv:2406.07610 [hep-th]} \BibitemShut {NoStop}%
\bibitem [{\citenamefont {Kristiano}\ and\ \citenamefont {Yokoyama}(2025)}]{Kristiano:2025ajj}%
  \BibitemOpen
  \bibfield  {author} {\bibinfo {author} {\bibfnamefont {J.}~\bibnamefont {Kristiano}}\ and\ \bibinfo {author} {\bibfnamefont {J.}~\bibnamefont {Yokoyama}},\ }\href@noop {} {\  (\bibinfo {year} {2025})},\ \Eprint {http://arxiv.org/abs/2504.18514} {arXiv:2504.18514 [hep-th]} \BibitemShut {NoStop}%
\bibitem [{\citenamefont {Inomata}(2025)}]{Inomata:2025bqw}%
  \BibitemOpen
  \bibfield  {author} {\bibinfo {author} {\bibfnamefont {K.}~\bibnamefont {Inomata}},\ }\href {\doibase 10.1103/PhysRevD.111.103504} {\bibfield  {journal} {\bibinfo  {journal} {Phys. Rev. D}\ }\textbf {\bibinfo {volume} {111}},\ \bibinfo {pages} {103504} (\bibinfo {year} {2025})},\ \Eprint {http://arxiv.org/abs/2502.08707} {arXiv:2502.08707 [astro-ph.CO]} \BibitemShut {NoStop}%
\end{thebibliography}%

\end{document}